\documentclass[sigconf]{acmart}

\AtBeginDocument{%
  }

\copyrightyear{2025}
\acmYear{2025}
\setcopyright{cc}
\setcctype{by-sa}
\acmConference[FAccT '25]{The 2025 ACM Conference on Fairness, Accountability, and Transparency}{June 23--26, 2025}{Athens, Greece}
\acmBooktitle{The 2025 ACM Conference on Fairness, Accountability, and Transparency (FAccT '25), June 23--26, 2025, Athens, Greece}\acmDOI{10.1145/3715275.3732140}
\acmISBN{979-8-4007-1482-5/2025/06}

\usepackage{shortcuts}
\usepackage{sankey}
\usepackage{subfigure}
\usepackage{subcaption}
\usepackage{booktabs,multirow}

\usepackage{array}
\usepackage{enumitem}
\usepackage{float}

\makeatletter
\renewcommand\paragraph{\@startsection{paragraph}{4}{\z@}%
  {3.25ex \@plus1ex \@minus.2ex}%
  {-0.3em}%
  {\normalfont\normalsize\bfseries}}
\makeatother

\usepackage{tabularx}

\begin{document}

%%
%% The "title" command has an optional parameter,fz
\title{Algorithms in the Stacks: Investigating automated, for-profit diversity audits in public libraries}
% \title{Algorithms in the Stacks: Investigating data-driven, for-profit diversity audits in public libraries}

%%\orcid{1234-5678-9012}
%\email{webmaster@marysville-ohio.com}
%\affiliation{%
%  \institution{Institute for Clarity in Documentation}
%  \country{USA}
%}

\newcommand{\uw}{$^\diamondsuit$}
\newcommand{\um}{$^\clubsuit$}
\newcommand{\sd}{$^\spadesuit$}
\newcommand{\aspace}{\hspace{\fontdimen2\font}}

\author{Melanie Walsh}
\orcid{0000-0002-1825-0097}
\email{melwalsh@uw.edu}
\affiliation{%
  \institution{University of Washington}
  \city{Seattle}
  \state{WA}
  \country{USA}
}

\author{Connor Franklin Rey}
\orcid{0000-0003-1234-5678}
\email{cfranklinrey@sdsu.edu}
\affiliation{%
  \institution{San Diego State University}
  \city{San Diego}
  \state{CA}
  \country{USA}
}

\author{Chang Ge}
\orcid{0000-0003-9876-5432}
\email{changgge@umich.edu}
\affiliation{%
  \institution{University of Michigan}
  \city{Ann Arbor}
  \state{MI}
  \country{USA}
}

\author{Tina Nowak}
\orcid{0000-0002-2468-1357}
\email{tnowak@uw.edu}
\affiliation{%
  \institution{University of Washington}
  \city{Seattle}
  \state{WA}
  \country{USA}
}

\author{Sabina Tomkins}
\orcid{0000-0001-1122-3344}
\email{stomkins@umich.edu}
\affiliation{%
  \institution{University of Michigan}
  \city{Ann Arbor}
  \state{MI}
  \country{USA}
}


\begin{abstract}
Algorithmic systems are increasingly being adopted by cultural heritage institutions like libraries.
In this study, we investigate U.S. public libraries' adoption of one specific automated tool---automated collection diversity audits---which we see as an illuminating case study for broader trends. 
Typically developed and sold by commercial book distributors, automated diversity audits aim to evaluate how well library collections reflect demographic and thematic diversity. We investigate how these audits function, whether library workers find them useful, and what is at stake when sensitive, normative decisions about representation are outsourced to automated commercial systems.
Our analysis draws on an anonymous survey of U.S. public librarians ($n=99$), interviews with 14 librarians, a sample of purchasing records, and vendor documentation. We find that many library workers view these tools as convenient, time-saving solutions for assessing and diversifying collections under real and increasing constraints.
Yet at the same time, the audits often flatten complex identities into standardized categories, fail to reflect local community needs, and further entrench libraries’ infrastructural dependence on vendors.
We conclude with recommendations for improving collection diversity audits and reflect on the broader implications for public libraries operating at the intersection of AI adoption, escalating anti-DEI backlash, and politically motivated defunding.
\end{abstract}




\maketitle

\begin{figure}[h!]
    \centering
    \includegraphics[width=\linewidth]{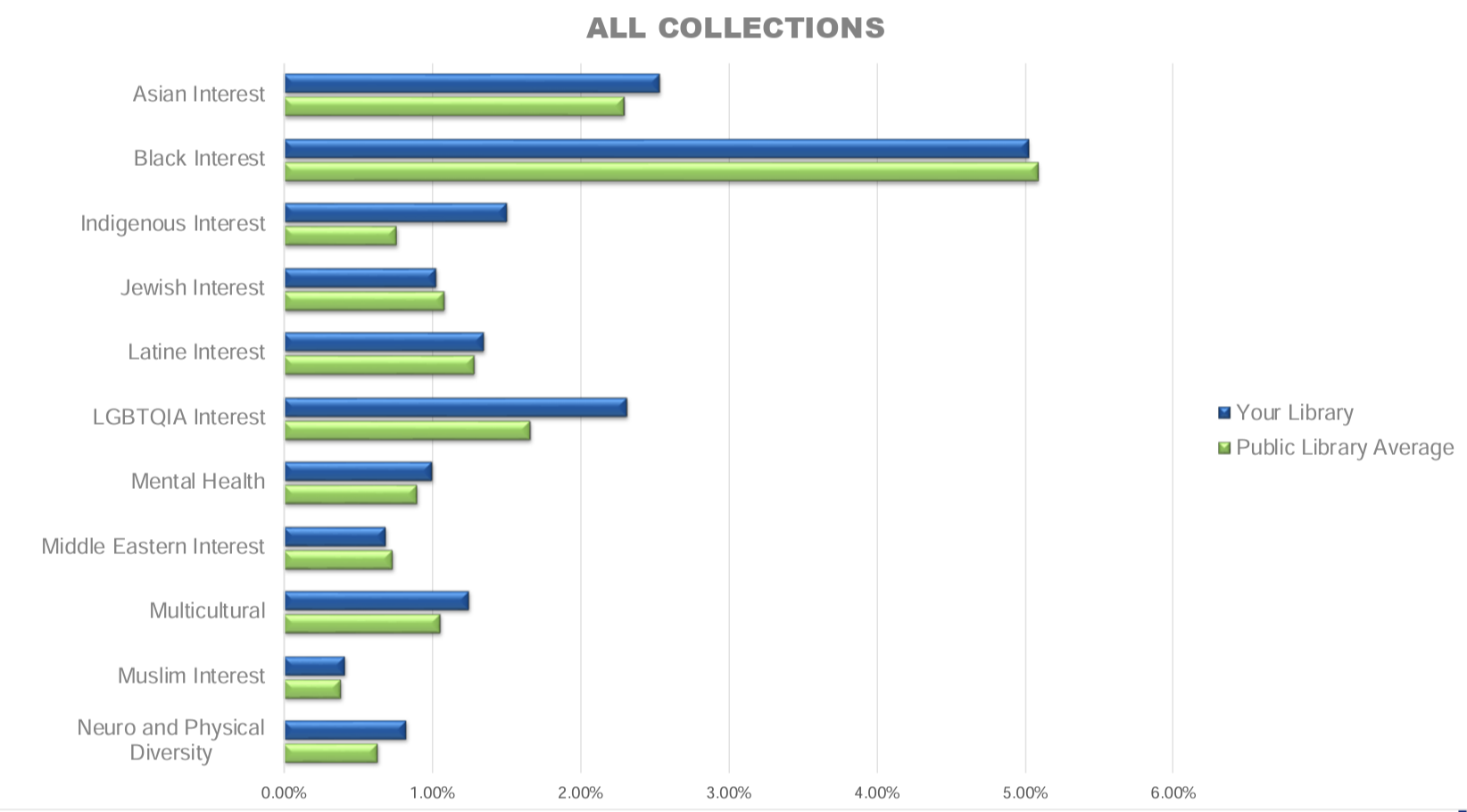}
    \caption{A library collection diversity audit report from the vendor Ingram, showing representation of books per diversity category.}
    \label{fig:bar-chart}
\end{figure}

\section{INTRODUCTION}

Algorithmic systems are increasingly being adopted by cultural heritage institutions like libraries, museums, and archives.
In U.S. libraries, these tools now include recommendation systems that personalize borrowing suggestions \cite{enisAIHorizon2024, priceNewBetaTest2023, OCLCIntroducesAIgenerated2023, malespinaAILibraryReaders2024}; chatbots that assist patrons with questions and search \cite{szydlowskiDrMartinLuther, GettingStartedPrimo2024, rodriguezLIBRARYCHATBOTSEasier2025, rodriguezUncodingLibraryChatbots2022}; predictive algorithms that forecast book circulation and checkout hold times \cite{enisBakerTaylorCollectionHQ2014, enisAIHorizon2024}; and automated tools that measure the ``diversity'' of thousands or even millions of library books (\figref{fig:bar-chart}). 
Like algorithmic systems adopted by other public institutions---such as courts, schools, and hospitals \cite{sendakHumanBodyBlack2020, freundDedicatedDocketUS2024}---these developments offer potential advantages and risks.
While automated tools are often embraced to perform tasks faster, at scale, and with a veneer of objectivity, extensive research at FAccT and elsewhere has shown that they can also reinforce and exacerbate biases, even when the goal is to root them out
% \citep{ab1,ab21,ab3,ab4,ab5,ab6,ab7}. 
\citep{obermeyerDissectingRacialBias2019, ab3, eubanksAutomatingInequalityHow2018}. 
% A similar phenomenon, with similar risks, is now emerging in , where algorithmic and data-driven tools are now being introduced. 

In this paper, we focus on the growing use of automated collection \textit{diversity audits} in public libraries---tools that quantitatively measure how well library holdings reflect demographic and thematic diversity.
% providing metrics for overall diverse representation, breakdowns in specific categories, and, typically, recommendations to fill the gaps. 
% We focus on large-scale tools developed and sold by third-party library vendors.
These tools are typically developed and sold by third-party vendors, and are rapidly becoming normalized in library practice.
While we focus on U.S. public libraries, collection diversity audits are increasingly being used in school and academic libraries, and outside the U.S. \cite{canadianschoollibrariesDiversityAudits2025, pedersenMeasuringCollectionDiversity2022}.
Broadly, we ask: \textbf{What happens when sensitive, value-driven cultural work is partly outsourced to automated systems developed by commercial vendors? Can these tools meaningfully support the values they claim to advance?}
% To ground this investigation, we focus on how automated diversity audits are being implemented, understood, and evaluated by public librarians. 

Central to these questions is the term \textit{diversity} itself. 
While at its most basic, diversity means difference within a group, the term has become widely used in U.S. institutional contexts to refer to structurally marginalized human identities and backgrounds, and especially Diversity, Equity, and Inclusion (DEI) efforts \cite{leongDiversityMessagingAffirmative2025a}. 
% DEI frameworks have drawn criticism from the right and left sides of the ideological spectrum---often for their ambiguity, which some argue is strategic.
We acknowledge the complexity and long history of the term \textit{diversity} in the U.S. \cite{ahmedBeingIncludedRacism2012, melamedDiversity2020, lawrenceEachOthersHarvest1997}.
For clarity, we broadly use \textit{diverse} and \textit{diversity} as they are deployed by vendors---to signal books that represent historically marginalized identities or subjects---and we examine how their systems define and operationalize those terms.
% We ask: Why have libraries turned to automated tools, especially those provided by third-party vendors, to measure diversity in books?
% What counts as a ``diverse'' book?
% And what might this case study suggest about the future of algorithms and AI in libraries and cultural heritage more broadly?

% Diversity audits of library collections are a recent but growing practice, developed in response to increasing awareness of systemic bias in libraries and the publishing industry \citep{So_Wezerek_2020,LibraryProfessionalsFactsa,alterPublishingPledgedDiversify2024a}.
% Historically, both fields are very white---and, despite some progress, they remain so today.
% While 60\% of the U.S. population identifies as white, more than 81 percent of librarians and more than 72 percent of publishing industry workers identified as white in 2023 \cite{LibraryProfessionalsFactsa, alterPublishingPledgedDiversify2024a}.
% The lack of diversity is even more pronounced in books themselves.
% More than 95\% of the books published between 1960 and 2018 by mainstream U.S. publishers were authored by white writers, according to a study by \citet{So_Wezerek_2020}. 
Collection diversity audits have emerged in response to longstanding inequities in both librarianship and book publishing \citep{So_Wezerek_2020,LibraryProfessionalsFactsa,alterPublishingPledgedDiversify2024a}.
While 61\% of the U.S. population identified as white in the most recent census, over 81\% of librarians and 72\% of publishing industry employees identified as white as of 2023 \cite{LibraryProfessionalsFactsa, lee&lowDiversityBaselineSurvey, bureau2020CensusIlluminates}.
The disparities are even greater in published books. 
Between 1950 and 2018, over 95\% of the most widely held titles released by mainstream U.S. publishers were authored by white writers, according to a study by \citet{So_Wezerek_2020}.
Because public libraries rely heavily on mainstream publishers, their collections are likely similarly skewed, limiting the kinds of stories and perspectives that are available to readers.
Diversity audits aim to assess and address these gaps.
They can prompt libraries to seek out more queer love stories; more narratives with Muslim protagonists; more books by and about Black and Indigenous people and other people of color.
% They help libraries benchmark, or measure, the representational makeup of their collections and identify areas where they need to acquire more books---for example, more books by and about Black, Indigenous, or other people of color; more narratives centering queer relationships; more stories featuring Muslim protagonists.
% The primary goal of a diversity audit is thus to help libraries benchmark their collections 
% % (that is, to take stock of representation in different categories)
% and to identify gaps where they need to acquire more books---

Yet in the increasingly hostile and reactionary political climate of the U.S.---marked by rising book bans and anti-DEI legislation \cite{ALA2024BannedBookData,thewhitehouseEndingRadicalWasteful2025, gardinerJudgeBlocksTrump2025, adamsMapSeeWhich2024, alfonsecaMapImpactAntiDEI2024,swaineAntiDEIPushNational2025, penamericaBannedUSAShelves2024}---diversity audits are not only used as tools to improve equity.
They have also been used defensively: to demonstrate that library collections are not ideologically skewed toward progressive agendas and to push back against censorship efforts \cite{schermeleCulturalPowerStruggle2022, escuderoHowAlbanyAdding2023}.
% ``The data shows the pushback is not founded in reality,'' one librarian told journalists after an audit. ``Our collections aren’t what they think they are.''  \cite{escuderoHowAlbanyAdding2023}
% Most early diversity audits were conducted manually or at a small scale, often with librarians physically pulling items from shelves, examining book covers, reading descriptions, and more. 

% While collection diversity audits emerged as a coherent trend in the 2010s, they began to grow in popularity in 2021 when vendors like \baker{}, Ingram, and OverDrive debuted larger, faster, and more convenient audits based on their own proprietary data and automated classification systems. 
While manual audits date back almost a decade \cite{jensenDoingYACollection2017}, the practice accelerated in 2021 when commercial vendors like \baker{}, Ingram, and OverDrive introduced large-scale audit tools powered by proprietary metadata and automated classification \cite{ingramIngramAnnouncesNew2021, symeAvailableNowAnalysis2021, staffGainGreaterInsight2022}. 
Where manually auditing just the children's section of a library might take months, these tools promise to audit entire collections in weeks or days.
The companies also provide ``shopping lists'' and recommended books to fill diversity gaps, which are typically books sold by the same vendor performing the audit.
% Despite their increased prevalence and popularity, little academic work has examined the significance or effectiveness of vendor-driven diversity audits \cite{voelsAuditingDiversityLibrary2022}. There is also limited FAccT scholarship exploring the impact of data and algorithmic tools on cultural heritage institutions like libraries, museums, and archives.

Despite their increased prevalence and popularity, little academic work has examined the effectiveness, accuracy, or implications of automated collection diversity audits \cite{voelsAuditingDiversityLibrary2022}---and few FAccT studies have explored algorithmic interventions in cultural heritage settings \cite{huangSocialInclusionCurated2022, joLessonsArchivesStrategies2020}.
To address these gaps, we draw on an anonymous survey of U.S. public librarians ($n$=99), in-depth interviews with 14 librarians, purchasing records from a sample of public libraries, and an analysis of vendor documentation and promotional materials.
We ask:
\begin{itemize} \item \textbf{RQ1:} How prevalent are automated collection diversity audits in libraries, and who provides them at what cost? \item \textbf{RQ2:} How do automated collection diversity audits actually work, and what counts as a diverse book? \item \textbf{RQ3:} Do library workers find these audits nuanced, trustworthy, and useful? \end{itemize}

We find that automated audits from commercial vendors are widely used and, for many library workers, offer a convenient tool for assessing collections under real and increasing constraints. 
Their appeal lies not only in their speed and scale, but also in their framing as neutral or objective---a framing that becomes especially valuable amid heightened political scrutiny of library content and DEI initiatives. 
Yet these tools often flatten the meaning of diversity, ignore structural drivers of inequality, and obscure the community-specific needs that shape thoughtful collection development. 
While the for-profit nature of these tools is not inherently problematic, we find that commercial incentives discourage methodological transparency, promote proprietary data hoarding, and deepen customer lock-in.
Book vendors now play a central role not only in supplying materials for collections but also in providing the data and analytics through which libraries interpret their collections.
This ``double-dip'' with data assets resembles information consolidation dynamics described by  \citet{lamdanDataCartelsCompanies2022}.
% While differing from larger information conglomerates, this expanding influence reflects broader infrastructural trends toward consolidation and opacity, as described by .
% These dynamics deserve scrutiny because even helpful and well-intentioned systems can have unintended consequences.
% Book vendors exert growing influence not just over materials, but also over the analytics infrastructure through which libraries interpret and understand their collections.
% This expanding role reflects similar dynamics to information monopolies described by \citet{lamdanDataCartelsCompanies2022}, even though they are not at the same scale.
% They also deepen libraries’ financial and technical dependence on external vendors and reflect a broader shift in these vendors’ roles---from suppliers of materials to providers of data analytics---echoing the consolidation of information infrastructure 

Ultimately, we suggest that library collection diversity audits are a critical case study for understanding the benefits and risks of using automated tools in cultural heritage settings, especially tools provided by commercial vendors.
While these tools do not currently (to our knowledge) incorporate Artificial Intelligence (AI) techniques, they exemplify the dynamics increasingly emerging with the application of AI to sensitive, subjective, and context-dependent cultural work.
We conclude with recommendations for improving these audits and reflect on the broader implications for libraries navigating the fraught intersection of AI adoption, escalating anti-DEI movements, and politically motivated defunding.

% We, as a varied group of researchers and library practitioners, believe that data and computational tools can be valuable for libraries, particularly as they manage large, complex collections and facilitate numerous interactions with patrons. We also believe diversity audits are important. But we argue that these tools, especially those developed and sold by vendors, deserve careful attention and scrutiny. 
% Though we remain critical of vendor-driven diversity audits in many respects, we also recognize that, in a political climate where libraries are actively under attack and underfunded, these audits often represent one of the few actionable tools available to address diversity and structural inequality. Now under a second Trump administration---with anti-DEI, anti-trans, anti-gay, and anti-immigrant legislation and sentiment intensifying---the sheer fact of conducting a diversity audit seems to feel like a meaningful political statement. We call attention to this difficult and complex climate for libraries.
\section{RELATED WORK}
% Three bodies of work are relevant to our study and discussed here. First, we review related audits of sociotechnical systems. Next, we discuss work on understanding how algorithms can exacerbate societal inequities. Finally, we discuss how race and identity has been measured in book collections. 

\subsection{Audits}
Audits, formal evaluations of systems, are a foundational concept in FAccT research, where they are most often applied to algorithmic and AI systems \cite{borradaileWhoseTweetsAre2020, mahomedAuditingGPTsContent2024, perreaultAlgorithmicMisjudgementGoogle2024, radiya-dixitSociotechnicalAuditAssessing2023}. 
In recent years, FAccT scholars have turned their attention not only to algorithms but to the audit process itself, interrogating its assumptions, politics, and institutional uses \cite{costanza-chockWhoAuditsAuditors2022, grovesAuditingWorkExploring2024, terzisLawEmergingPolitical2024}.
We draw on both strands of this work because collection diversity audits operate as automated tools and audits.
% , we draw on work related to the examination of algorithmic systems and the critical study of auditing practices.
% These tools promise to quantify the “diversity” of library collections using proprietary datasets and classification schemes. While vendor documentation often emphasizes objectivity and neutrality, we interrogate the social assumptions, classification logics, and commercial infrastructures embedded in these tools. 
In the field of Library and Information Science (LIS), scholars and librarians have begun to document and analyze collection diversity audits \cite{fuller-gregoryAuditsWholePicture, gatesUsingDataCollection2022, fischerAutomatingDiversityAudit2023, waltersAssessingDiversityAcademic2023, smithBuildingDiverseCollections2023a, voelsAuditingDiversityLibrary2022}.
We build on their work with a systematic analysis of how these tools operate in practice.

\subsection{Race, Diversity, and Algorithmic Classification}
A substantial body of research has shown that technology and algorithmic systems often replicate and exacerbate social inequalities, particularly along lines of race, gender, and other axes of identity \cite{benjaminRaceTechnologyAbolitionist2019, chunIntroductionRaceTechnology2009, nobleAlgorithmsOppressionHow2018, buolamwiniGenderShadesIntersectional2018}. 
These harms are not always the result of explicitly encoded categories but often emerge through data, design decisions, and institutional assumptions.

In the field of algorithmic fairness, race and social identity are often explicitly operationalized—that is, converted into measurable categories—in order to evaluate whether systems produce discriminatory outcomes. 
While this is frequently done in the name of justice and anti-discrimination, the process of categorizing and quantitatively measuring social identity can reproduce harmful histories of classification, especially when underlying assumptions go unstated and unexamined \cite{abduEmpiricalAnalysisRacial2023a}.
% that, across algorithmic fairness research, race is defined inconsistently, the rationales behind these definitions are rarely explained, and categories are often shaped by technical and computational priorities (e.g. using fewer categories makes measurement easier).
Further, \citet{hannaCriticalRaceMethodology2020} argue that the operationalization of race often obscures structural causes of discrimination: “treating race as an attribute, rather than a structural, institutional, and relational phenomenon...serves to minimize the structural aspects of algorithmic unfairness.''

For decades, feminist and critical race theorists have also interrogated the institutional uses of \textit{diversity}, sometimes critiquing its ambiguity, depoliticization, and neoliberal cooption, and sometimes acknowledging its strategic utility \cite{ahmedBeingIncludedRacism2012, melamedDiversity2020, lawrenceEachOthersHarvest1997, melamedRepresentDestroyRationalizing2011}.
These critiques resonate with concerns about diversity in recommendation systems, in which scholars have questioned how diversity is abstractly operationalized and argued that it must be defined in relation to the specific goals and contexts of each system \cite{vrijenhoekDiversityWhatDifferent2024}.
Scholarship from algorithmic fairness, recommendation systems, and feminist and critical race theory thus highlights problems and complexities inherent in operationalizing race and diversity as collection diversity audits do.

\subsection{Measuring Race and Identity in Books}
Scholars in the digital humanities (DH) and LIS have long grappled with the challenges of categorizing race and other aspects of social identity in books, especially with computational methods or metadata.
In DH, scholars have emphasized that computation and racist discrimination are historically and structurally entangled \cite{gallonMakingCaseBlack2016, mcphersonWhyAreDigital2012}.
% Research in the digital humanities (DH) has pointed out that computation and racist discrimination are intimately entangled \cite{gallonMakingCaseBlack2016, mcphersonWhyAreDigital2012}, and the
While researchers have used computational approaches to explore race, gender, and social identity in literary and cultural texts \cite{soRaceDistantReading2020, changQueerGapCultural2023, changSubversiveCharactersStereotyping2024}, many foreground the myriad challenges involved in doing so \cite{mandellGenderCulturalAnalytics2019}.
In fiction, characters' identities are not always stated outright; they can be fluid or ambiguous; they can be contingent on historical period, geography, or context.
Much of this is also true for real-life authors \cite{soRaceDistantReading2020}.
Researching an author's background is not only time-consuming but often leads to findings that defy neat categorization---as \citet{soRaceDistantReading2020} point out, the American author Nella Larsen is typically identified as Black today but in the 1920s (when she was writing) was often referred to as ``mulatta,'' a term that is outdated, offensive, and reflective of a different racial classification system.  

Within LIS scholarship, similar concerns have been raised about the classification of race, ethnicity, gender, sexuality, and Indigeneity through metadata like \textit{subject headings}. 
In particular, scholars have critiqued the U.S. Library of Congress Subject Headings (LCSH) for outdated, inconsistent, or offensive treatment of marginalized identities \cite{duarteImaginingCreatingSpaces2015, baronChangeSubject2019, watsonHomosaurusDigitalTransgender2019, zwaafHomosaurusHttpHomosaurusorg2020a, littletreeKnowledgeOrganizationIndigenous2015, billeyInclusiveCatalogingHistories2024, drabinskiQueeringCatalogQueer2013, cataloginglabProblemLCSH2025, bermanPrejudicesAntipathiesTract2013}. 
Our work builds on these DH and LIS traditions to critically examine how diversity categories are constructed and measured with collection audit tools.

\section{BACKGROUND}
% Our work is informed by the study of 
% library purchasing, and how such purchasing has become ``data-driven''. Finally we discuss previous work on understanding the subject of our study, diversity audits, a particular case of library purchasing which is increasingly sold as an algorithmic purchase. 
\subsection{Major Library Vendors in the U.S.}
Public libraries rely on a range of commercial vendors for everything from furniture and security systems to digital platforms and book distribution. 
This study focuses on three major players in the public library market---Baker \& Taylor, Ingram, and OverDrive---that sell books as well as automated and data-driven tools, including diversity audit tools.
Companies like Midwest Tape and LibraryIQ also offer tools for collection diversity audits \cite{MidwestTapeBlack2022, libraryiqDiversity2025}.
Some free, nonprofit tools exist, as well, such as \href{https://diversebookfinder.org/}{Diverse BookFinder}, which focuses on picture books, and a tool from the \href{https://www.ccslib.org/}{Cooperative Computer Services} (CCC) Consortium \cite{fischerAutomatingDiversityAudit2023}.
We focus on these three vendors because they are widely used and exemplify the broad entanglement of book distribution, metadata infrastructure, and automated systems.

\begin{itemize}[leftmargin=*, itemsep=0.3em]
    \item \textbf{Baker \& Taylor} (founded 1828): The largest U.S. library supplier, offering book distribution, cataloging, metadata, and collection development services. Acquired \textit{collectionHQ} in 2011 to provide data-driven collection analysis and algorithmic tools.
    
    \item \textbf{Ingram} (founded 1970): The largest overall U.S. book distributor, serving both libraries and retailers like Barnes \& Noble. Offers cataloging, processing, and AI-driven tools. Owns the self-publishing platform \href{https://www.ingramspark.com/}{IngramSpark}.
    
    \item \textbf{OverDrive} (founded 1986): The leading distributor of digital content (ebooks, audiobooks) to libraries. Operates the Libby app and videostreaming app Kanopy. Serves 95\%+ of U.S. public libraries \cite{OverDrivesNewOwners}.
    
\end{itemize}

\subsection{From Book Distribution to Data Analytics}

The role of commercial library vendors like book distributors has expanded significantly over time---particularly with the rise of digital materials and computing technologies, and the decline of public funding for libraries.
Rather than buying books directly from publishers, libraries have long relied on wholesalers like Baker \& Taylor (a company that has existed for almost 200 years), to supply books in bulk.
When catalog systems were computerized in the 1970s and 1980s---replacing physical card catalogs that listed each book's title, author, subjects, and location in the library---vendors gained a new foothold because they could ship books with machine-readable metadata (known as MARC records), allowing libraries to automatically populate catalog entries at scale \cite{avramMARCItsHistory1975, kilgourHistoryLibraryComputerization1970, rushLibraryAutomationSystems1982, rushLibraryAutomationMarket1988, matthewsAutomatedCirculationMarketplace1982}. 
% Library vendors, especially book distributors, have played a pivotal role in the functioning of public libraries for nearly two centuries. 

Vendors' early foothold in metadata allowed them to expand further into the business of data and algorithmic tools in the twenty-first century---especially as collections have grown and become more digital, and as libraries have faced staffing shortages, technical skill gaps, and funding issues \cite{kempMARCRecordServices2008, martinCatalogingEBooksVendor2010, collinsCurrentBudgetEnvironment2012}. 
In 2011, for example, \baker{} acquired the Scottish startup product \textit{collectionHQ}, which helps libraries manage their collections by analyzing borrowing patterns and predicting future circulation with machine learning. 
These tools are used both to guide new book acquisitions and to recommend which titles should be “weeded,” or removed, from the collection—a core task for libraries with limited shelf space and budgets.

% This saved libraries considerable time and labor, and further entrenched vendors’ role in daily operations.

% The shift began in earnest in the 1970s and 1980s when cataloging systems were first computerized. 
% Previously, every library book was cataloged by hand using physical cards, which listed information such as the title, author, and subjects. 
% With the introduction of Integrated Library Systems (ILS), vendors gained a new foothold: they could now ship books with machine-readable metadata, allowing libraries to automatically populate catalog entries. 
% This saved libraries considerable time and labor, and further entrenched vendors’ role in daily operations.

These developments resemble the broader trajectory of information conglomerates like RELX and Thomson Reuters, which have acquired more and more information companies (e.g., Elsevier, Westlaw), and which can “double-dip” with their massive data assets—selling both raw data and structured analytics derived from that data \cite{lamdanDataCartelsCompanies2022}. 
While vendors like Baker \& Taylor differ from these conglomerates in many ways, including the entities they contract with, they too have acquired more information companies (e.g., collectionHQ) and more library data, and now sell back analytics and tools built from this aggregated data.

% The introduction of computerized cataloging systems, or Integrated Library Systems (ILS), in the 1970s and 1980s further embedded vendors into library operations, as they began providing pre-cataloged materials and metadata tailored to these systems. 
% This integration saved libraries considerable time and resources, solidifying vendors’ position in the library ecosystem.
% In the twenty-first century, vendors moved into the business of data and increasingly algorithms and AI. 
% In 2011, \baker{} acquired the Scottish start-up \textit{collectionHQ}, which leverages data and algorithmic tools to help libraries manage their collections. 
% For example, by predicting which books will circulate in a given community, these tools help point libraries toward what titles to buy, and also help target books that should be ``weeded'' or removed from the collection (a fundamental and ongoing task for libraries with finite shelf space and resources).

% These developments mirror the trajectory of information conglomerates like RELX and Thomson Reuters, who \cite{lamdanDataCartelsCompanies2022} argues can ``'double-dip' with their data assets, selling raw data, and also selling structured information made from that raw data.''
% In a similar vein, public libraries pay vendors to manage their collection data, and, increasingly, vendors sell back various structured versions of that data.

\subsection{Collection Diversity Audits}

Library collection diversity audits emerged as a coherent trend in the 2010s with predominantly manual processes \citep{voelsAuditingDiversityLibrary2022, jensenDiversityAuditing101, jensenDoingYACollection2017}. 
% Karen Jensen, a librarian and writer for the \textit{Teen Librarian Toolbox} blog, is credited with one of the earliest documented diversity audits in 2016 \citep{voelsAuditingDiversityLibrary2022, jensenDiversityAuditing101, jensenDoingYACollection2017}. 
Of course, attention to the lack of diverse books in library collections and the publishing industry pre-dates the 2010s \cite{chadley1992addressing, larrickAllwhiteWorldChildrens1965}.
% Of course, attention to the lack of diverse books offered by the publishing industry and calls for diversity in library collections pre-date the 2010s \cite{chadley1992addressing}. 
Many librarians have long been conscientious about developing collections that offer ``mirrors, windows, and sliding glass doors'' for their communities \citep{bishopMirrorsWindowsSlidingGlassDoros}. 
% \citet{larrickAllwhiteWorldChildrens1965} is even known to have conducted an assessment akin to a diversity audit as early as 1965. 
However, in the past, librarians mostly lacked formalized processes to measure the diversity of their collections comprehensively. 

\subsubsection{How Do (Manual) Collection Diversity Audits Work?}

There are many documented instances of libraries conducting manual audits, including at public, academic, and school libraries \citep{ciszek2010diversity, EMERSON2022102517, mortensen2019measuring, sevits2024developing}. 
% The methods used vary between libraries. 
In a manual collection audit, librarians typically define a list of diversity categories---often related to race, ethnicity, religion, disability, and/or LGBTQ+ representation.
To measure how many books are in these categories, they often sample sections of the collection and physically handle the books, evaluating book covers, illustrations, publication dates, subject headings, and sometimes reading excerpts.
Auditors may additionally consult book reviews, prize lists, or author biographies---although the identification of \#OwnVoices titles (where author and protagonist share a minoritized identity) has faced criticism due to privacy concerns \citep{voelsAuditingDiversityLibrary2022, jensenDiversityAuditing101, jensenDoingYACollection2017}.

% Some libraries also collect data on any identity groups authors may belong to, in order to evaluate the diversity of authors and \#OwnVoices titles (books in which the author and protagonist share a minoritized identity). However many avoid collecting this data because of the workload involved in searching for this information, as well concerns over author privacy 

 % Manual audits often focus on a specific subset of a collection---e.g., Children's Picture Books, Teen Fiction, or Adult Non-Fiction---and use methods including examining book covers, researching authors' backgrounds, consulting book reviews or prize lists, or noting protagonists' identities \citep{rothbauer1999gay,williams2014diverse, kristick2020diversity}. 

\subsubsection{Rise of Automated Vendor Audits}
In 2020, collection diversity audits became more prevalent in the U.S.
Responding to the unjust murder of George Floyd, the 2020 Black Lives Matter (BLM) protests inspired organizations around the U.S. to invest in various DEI and anti-racism initiatives.
Many interviewees reported that the confluence of BLM protests and the COVID-19 pandemic---which forced many libraries to close to the public and thus loosened some operational capacity---was a key reason they began a diversity audit. 
In summer 2021, Ingram launched a diversity audit service called \textit{iCurate inClusive} \cite{ingramIngramAnnouncesNew2021}, and Baker \& Taylor introduced a \textit{DEI Analysis} tool through their subsidiary, \textit{collectionHQ} \cite{symeAvailableNowAnalysis2021}. 
Soon after, other vendors like OverDrive \cite{staffGainGreaterInsight2022}, Midwest Tape (which owns Hoopla, provider of ebooks and digital content) \cite{MidwestTapeBlack2022}, Follet \cite{staffUsingFollettTitlewave2022}, and Mackin \cite{mackinTagReportAnalyze2025} also developed diversity audit/analysis services.
% According to vendors we spoke with, librarians also called on vendors to develop new diversity audit tools around this time.

% These calls, combined with the time- and labor-intensive nature of manual audits, prompted vendors to develop large-scale diversity audit tools based on their own proprietary data and automatic classification systems. 

% The number of U.S. libraries that have conducted a diversity audit has risen rapidly in the last five years. According to \textit{Library Journal}, the percentage of public libraries that had conducted a collection diversity audit rose from 5.5\% in 2019 to 46\% in 2022 \cite{vercellettoHowDiverseAre,wyattCollectionRebalance2022a}.

% \vspace{-1em}
\subsection{Book Classification and Subject Headings}

Libraries and vendors primarily rely on established classification systems to categorize books, and these systems play a central role in automated diversity audits.
% While classification might involve many types of metadata, in the context of libraries and publishing, 
Standardized topical labels—often called \textit{subject headings}—are typically assigned when a book is entered into a library system, and serve as a key way to indicate what a book is “about.”

% Subject headings play a fundamental role in library systems, and without them, it would be nearly impossible to organize, retrieve, or analyze the topical content of books at scale. 
% They serve as the metadata backbone of library catalogs and are essential for search, shelving, and discovery.

Two of the most commonly used systems are \textbf{Library of Congress Subject Headings (LCSH)} and \textbf{Book Industry Standards and Communications (BISAC)} codes.
LCSH, maintained by the U.S. Library of Congress, is a controlled vocabulary that includes topics, geographies, time periods, genres/forms, ethnic groups, and more.
While LCSH is widely used, there have been ongoing criticisms of its treatment of race, ethnicity, gender, sexuality, and Indigeneity, and substantive change has been slow \cite{duarteImaginingCreatingSpaces2015, baronChangeSubject2019, watsonHomosaurusDigitalTransgender2019, zwaafHomosaurusHttpHomosaurusorg2020a, littletreeKnowledgeOrganizationIndigenous2015, billeyInclusiveCatalogingHistories2024, drabinskiQueeringCatalogQueer2013, cataloginglabProblemLCSH2025}.
BISAC headings, developed by the Book Industry Study Group (BISG), are used primarily in commercial publishing and bookselling, though some libraries use this system, too. 

% \vspace{-1em} % Reduce space before the table
\begin{table}[ht!]
\centering
\footnotesize
\setlength{\tabcolsep}{2pt}
\renewcommand{\arraystretch}{1.05}
\begin{tabular}{>{\raggedright\arraybackslash}p{1.4cm} >{\raggedright\arraybackslash}p{3.2cm} >{\raggedright\arraybackslash}p{3.2cm}}
\toprule
\textbf{Title} & \textbf{Library of Congress Subject Headings (LCSH)} & \textbf{Book Industry Standards and Communications (BISAC) Subject Headings} \\
\midrule
\textit{Fun Home} (2006) \par Alison Bechdel 
& \textbullet{} Bechdel, Alison, 1960– — Comic books, strips, etc. \par
  \textbullet{} Cartoonists — United States — Comic books, strips, etc. \par
  \textbullet{} Graphic novels \par
  \textbullet{} Comics (Graphic works) \par
  \textbullet{} Nonfiction comics
& \textbullet{} COMICS \& GRAPHIC NOVELS: Nonfiction / Biography \& Memoir \par
  \textbullet{} COMICS \& GRAPHIC NOVELS: LGBTQ+ / General \par
  \textbullet{} COMICS \& GRAPHIC NOVELS: Humorous \\
\midrule
\textit{Beloved} (1987) \par Toni Morrison 
& \textbullet{} African Americans -- Ohio -- History -- 19th century -- Fiction \par
  \textbullet{} African Americans -- Social conditions -- To 1964 -- Fiction \par
  \textbullet{} African American women -- Fiction \par
  \textbullet{} Enslaved women -- Fiction \par
  \textbullet{} Slavery -- United States -- Fiction \par
  \textbullet{} Infanticide -- Fiction \par
  \textbullet{} Ohio -- Fiction \par
  \textbullet{} Post-Civil War
& \textbullet{} Fiction / Literary \par
  \textbullet{} Literary Fiction \par
  \textbullet{} Popular Fiction \\
\midrule
\textit{The Satanic Verses} (1988) \par Salman Rushdie 
& \textbullet{} Airplane crash survival -- Fiction \par
  \textbullet{} Life change events -- Fiction \par
  \textbullet{} East Indians -- England -- Fiction \par
  \textbullet{} London (England) -- Fiction
& \textbullet{} Fiction / Literary \par
  \textbullet{} Fiction / Fantasy / Paranormal \par
  \textbullet{} Fiction / Psychological \par
  \textbullet{} Popular Fiction \\
\bottomrule
\end{tabular}
\caption{Subject headings for three well-known fiction titles. The Library of Congress Subject Headings (LCSH) are drawn from the Seattle Public Library, and the \textbf{Book Industry Standards and Communications (BISAC)} subject codes are drawn from publishers' websites.}
\label{tab:fiction_subjects_singlecol}
\end{table}
% \vspace{-2em} % Reduce space after the table

Most libraries do not assign subject headings from scratch but instead receive cataloging records from vendors or central services like \href{https://www.oclc.org/en/home.html}{OCLC}.
% which contain pre-populated subject headings. 
% OCLC is not pre-populated; bibliographic data which includes subject may be defined; OCLC owns Dewey
% Most libraries use Deweytope
Publishers assign BISAC headings during production.
% , and these codes may be imported directly into library systems. 
Catalogers at each library then decide whether to keep, modify, or supplement these headings based on local policies and community needs. 
% Large libraries may have dedicated cataloging staff, while 
Smaller libraries often rely heavily on vendor-supplied metadata. 

To illustrate how these systems operate in practice, Table~\ref{tab:fiction_subjects_singlecol} presents subject headings for three well-known fiction titles.
% , drawn from the Seattle Public Library (in turn likely drawing on vendor or OCLC records) and publisher websites. 
These examples show that subject headings vary in granularity and emphasis across classification systems—and that critical themes may be inconsistently labeled or entirely absent depending on the source, such as the subject of Islam in Salman Rushdie's \textit{The Satanic Verses} (1988), slavery in Toni Morrison's \textit{Beloved} (1987), or queerness in Alison Bechdel's \textit{Fun Home} (2006).
Because vendor diversity audits often rely on subject headings, their effectiveness depends on how books are cataloged and whether the metadata accurately reflects the thematic and demographic dimensions under evaluation.

\section{METHODS}
% We employ a mixed-methods approach. 
% To understand how library practitioners view collection diversity, library purchases, and diversity audits, we interview public library workers from across the United States. 
% To quantify library purchasing, we draw on public records of library spending. 

\subsection{Interviews}

We interviewed 14 staff members from public libraries across the United States, including collection development librarians, library directors, and library assistants (see more details in Appendix \ref{appendix-interviews}). 
We conducted five initial interviews that helped inform our later survey (two in July 2023, three in late 2024–early 2025). 
Then, in January 2025, seven survey participants elected to participate in follow-up interviews, with two additional interviewees identified via snowball sampling. 

All interviews were conducted virtually using video conferencing software and transcribed through speech-to-text software. 
The first, second, and fourth authors conducted an iterative thematic analysis \cite{clarkeThematicAnalysis2017} of interview transcripts using Atlas.ti. 
We began with an open coding phase, collaboratively identifying emergent themes. 
Through ongoing discussion and refinement, we consolidated tags into a smaller, structured set of themes. 

\subsection{Survey}

We developed a survey to evaluate librarians' experiences with and perspectives on diversity audits with a focus on public libraries. 
We distributed the survey to relevant mailing lists and online communities for public libraries and collection management (e.g., American Library Association Connect; mailing lists for the Association of Rural and Small Libraries, Urban Libraries Council, etc.). 
We received 99 quality responses from public library workers; 90 respondents filled out the entire survey, including basic demographic information about their library (see more details in Appendix \ref{appendix-survey}).

\begin{table}[!htb]
    \parbox{.4\linewidth}{
      \centering \small
        \begin{tabular}{lc}
        \toprule
        \textbf{Region}   & \textbf{Respondents} \\ 
        \midrule
        Northeast         & 37 (45\%)                \\ 
        West              & 21 (26\%)                \\ 
        Midwest           & 14 (17\%)                \\ 
        Southeast         & 8 (10\%)                \\ 
        Southwest         & 2 (2\%)                 \\ \bottomrule
         \multicolumn{2}{c}{(a)}  \\
        \end{tabular}
        \label{tab:region_percentages}
    }
    \hfill
    \parbox{.54\linewidth}{
      \centering \small
       \begin{tabular}{lc}
        \toprule
        \textbf{Library Pop. Size}         & \textbf{Respondents} \\ \midrule
        Fewer than 1,000             & 2 (2\%)                     \\ 
        1,000–4,999                  & 2 (2\%)                      \\ 
        5,000–9,999                  & 6 (7\%)                      \\
        10,000–24,999                & 14 (16\%)                     \\
        25,000–49,999                & 19 (21\%)                     \\ 
        50,000–99,999                & 17 (19\%)                     \\
        100,000–249,999              & 15 (17\%)                    \\
        250,000–499,999              & 9 (10\%)                    \\ 
        500,000 or more              & 6 (7\%)                      \\ \bottomrule
        \multicolumn{2}{c}{(b)}  \\
        \end{tabular}
        \label{tab:library_sizes_sorted}
    }
    \caption{(a): Public library survey respondents by region of library. (b): Public library survey respondents by library population size. (\( n = 90 \))}
\end{table}

\subsection{Purchasing Data}

We obtained public library purchasing records from a platform called GovSpend\footnote{https://govspend.com/govspend-platform/}, which tracks federal, state, and local government procurement for over 8.8 million entities, including many public libraries.
GovSpend is typically used by contractors and government entities to understand the market for their services or historical contract costs, but it has also been used to study government spending more broadly
\cite{corrigan2022banned}.

\subsubsection{GovSpend Search}
We query the GovSpend database for records where the item price is over \$20, the agency name contains ``library,'' and the description includes the words ``diversity,'' ``DEI,'' or the specific product names ``iCurate'' and ``collectionHQ.''
We manually review the results to exclude irrelevant items (e.g., ``pencils in \textit{diverse} colors'').
This process yields 171 entries broadly related to diversity initiatives, including 35 that directly pertain to collection diversity audits.
% We organize the records into two groups: \textbf{broad DEI-related purchases}, which contain 171 entries that we aggregate and display in Figure \ref{fig:sub2}; \textbf{focused collection diversity audit purchases}, which contain 35 purchases that we aggregate and represent in Figure \ref{fig:sub1}. 

% These 35 purchases were from 24 libraries across 13 states. 
% into our research scope. Approximately  purchase records match our search. 

% \begin{itemize}
%     \item \textbf{Broad DEI-related purchases}. In our broadest category, we include items that are meaningfully related to diversity and DEI in libraries, such as ``staff day DEI training,'' ``inclusive audit service,'' and ``collection diversity,'' and we exclude items that are not (e.g., purchases of office supplies like a ``new DEI director chair'' or ``pencils in diverse colors'', or natural science books with titles of `biodiversity` and ``diverse metabolic process''). This screening process yielded a result of 171 entries, which we aggregate and display in Figure \ref{fig:sub2}. These 171 purchases were from 57 libraries across 21 states. 
%     \item \textbf{Focused collection diversity audit purchases}. From these 171 entries, we further zoomed in on 35 entries that were directly related to collection diversity audit services, which we aggregate and represent in Figure \ref{fig:sub1}. These 35 purchases were from 24 libraries across 13 states. 
% \end{itemize}

To contextualize diversity-related purchases within broader library spending, we collect records over \$100 for two large library systems within multi-year time frames (2021-2024).
Library A serves a Midwestern city with a population exceeding 900K, including approximately 650K active library cardholders. Library B is located in another Midwestern city with a population of around 300K, with over 100K active cardholders.
% We summarize their purchases from companies of interest in Table \ref{tab:LibC} and Table \ref{tab:LibL}.

% Library C purchased a diversity audit from Ingram (iCurate inClusive service, on average \$3,600 per year). Library L purchased a DEI collection tool (\$223.74) from Baker \& Taylor. 

% The library reports an annual circulation of 13.1 million items, with 4 million eContent downloads. It receives approximately 5 million in-person visits and 9 million web visits each year. 

% The library attracts more than 2.3 million visits annually and boasts a collection of over 4 million books and other items. 
\subsection{Vendor Materials and Background Research}

To better understand how automated collection audits work, we also review available vendor documentation, including product descriptions, marketing materials, webinars, and case studies. 
Additionally, we conduct informal background interviews with vendor representatives. 
% ]to clarify system functionality and underlying assumptions.
% This analysis helped inform our interpretation of how automated audits are designed and implemented.

\subsection{Limitations}

Multiple considerations limit the extent to which our findings can be generalized.
% These three sources capture perspectives from librarians and libraries across the US, including both major cities and rural communities.
% However, each data source has limitations. 
The GovSpend platform does not include purchasing records for all public libraries, and its data---collected through public records requests and web scraping---may be incomplete or inconsistently labeled. 
Our survey was distributed through professional networks and online communities, resulting in a self-selecting and self-reported sample. 
Because it was anonymous, we also cannot rule out the possibility that multiple respondents came from the same institution. 
Nonetheless, responses span 26 states and 55 unique state/library population size combinations, indicating substantial geographic and institutional diversity.
% Additionally, the interview sample is relatively small and not fully representative of public librarians nationwide. 
Several interviewees were identified because they or their libraries had publicly discussed conducting diversity audits, which may bias the sample toward more engaged perspectives.
% We identified several interviewees because they or their library had publicized their engagement with diversity audits.
Lastly, because our study was conducted before Donald Trump took office as president in 2025 and before the latest wave of anti-DEI legislation, we do not capture the most recent political attacks and institutional constraints now being imposed on libraries.
Future work in this area is essential.

\begin{table*}[ht]
\centering
\setlength{\tabcolsep}{6pt} 
\renewcommand{\arraystretch}{1.2}
\footnotesize
\begin{tabular}{p{2.5cm}p{4.5cm}p{4.5cm}p{4.5cm}}
\toprule
\textbf{Features} & \textbf{Baker \& Taylor} & \textbf{Ingram} & \textbf{OverDrive} \\
\midrule
% \textbf{Tool Name} 
% & DEI Analysis
% & iCurate inClusive
% & Diversity Audit \\
\textbf{Classification Method \& Sources} 
& Automated matching on: subject headings (LCSH, BISAC), proprietary lists of books mentioned in library media sources or identified by staff; user self-classification (some crowdsourcing).
& Automated matching on: subject headings (LCSH, BISAC), proprietary lists of books identified by staff (curated for 20+ years); manual staff review.
& Automated matching on: subject headings (BISAC), lists of books that won diversity awards (e.g., Stonewall Book Awards, Transgender Voices). \\
\textbf{Audit Scope} 
& Books sold by any vendor; any language (as long as it matches classification criteria).
& English-language print books sold by Ingram (19+ million titles); includes some out-of-print books.
& Digital books sold by OverDrive; any language (as long as it matches classification criteria).\\
\textbf{Recommendations} 
& Purchasable recommended titles.
& ``Shopping lists'' of recommended Ingram books ranked by library sales popularity.
& Recommended OverDrive titles. \\
\textbf{Speed}
& Monthly updated results.
& 2 weeks.
& At least 5 days. \\
\textbf{Price} 
& Free with collectionHQ subscription; \newline \$6-13,500+ for standalone for first year.
& \$1,500 (single age group); \$4,200 (all age groups); Additional fee for follow-up audit.
& Free with OverDrive subscription. \\
\textbf{Reports} 
& Interactive dashboard with gap analysis; exportable CSVs with identified diverse books.
& Spreadsheet results; Slideshow presentation; Peer (average public library) benchmarking.
& Live presentation; PDF report; Supplemental materials including all BISAC headings used. \\
\bottomrule
\end{tabular}
\caption{Comparison of vendor-provided diversity audit tools. This information is drawn from public vendor documentation and background interviews with vendor staff \cite{baker&taylorCollectivePurchasingAgreements2025,baker&taylorIntroducingBakerTaylors2022, ingramICurate2025, ingramICurateInClusiveSchools2025, overdriveDiversityAuditsOverDrive2025, staffGainGreaterInsight2022, collectionhqDiversityEquityInclusion2025, bakerandtaylorBakerTaylorDiversity2025}.}
\label{tab:vendor-tools}
\end{table*}

\section{KEY FINDINGS}
% We highlight six key findings. Several of our findings characterize the relationship between libraries and vendors. For example, we find that diversity audits can enable vendor dependency. We also discuss the consequences of these audits, e.g, that they are viewed as homogenizing content. Finally, we share how the convenience they offer induces libraries to choose to conduct vendor-driven audits, despite their drawbacks.
% We present key findings in response to our three research questions: 
% \begin{itemize}
%   \item \textbf{RQ1:} How prevalent are automated diversity audits in libraries, and who provides them at what cost?
%   \item \textbf{RQ2:} How do automated diversity audits actually work?
%   \item \textbf{RQ3:} Do library workers find these audits nuanced, trustworthy, and useful?
% \end{itemize}
% We organize our findings thematically but note where each insight responds to one or more of these guiding questions.

% \subsection{How Do Automated Diversity Audits Work? What Counts as a ``Diverse'' Book? (RQ2)}

\subsection{Vendor Audits Use Standardized Diversity Categories and Proprietary Metadata (RQ2)}
\label{categories}

% To examine how automated collection diversity audits actually work (RQ2), we analyze vendor documentation, background interviews with vendor representatives, and librarians' accounts from our surveys and interviews. 
We find that vendors operationalize diverse books by devising standardized categories, and then classifying books by automatically matching on a combination of subject headings and proprietary data sources, such as title lists curated by vendor staff (Table \ref{tab:vendor-tools}).
Specific diversity categories vary slightly by vendor, but they all follow a similar structure and exhibit a degree of conceptual incongruity (Table~\ref{tab:diversity-categories}). 
Vendors combine demographic markers (e.g., ``Black'', ``Asian Interest''), thematic social issues (e.g., ``Substance Abuse \& Addictions''), and umbrella terms (e.g., ``Multicultural''), resulting in an expansive but loosely organized schema.
Many library workers raised concerns about the breadth and ambiguity of these frameworks (Section~\ref{homogenize}). 
% In particular, librarians questioned whether such generalized metrics could meaningfully reflect the complex identities and needs of the communities they serve.

% All three vendors include categories related to race and ethnicity, such as “Asian,” “Black,” and “Hispanic \& Latino.” They also commonly include gender, sexuality, disability, mental health, and religion. But the category systems expand beyond demographic labels to include topics like “Substance Abuse \& Addiction,” “Alternative Family,” “Immigration,” and “Equity \& Social Issues.”

To assign books to categories, vendors typically draw on a library's catalog metadata and computationally match book identifiers---such as ISBNs or EANs\footnote{The International Standard Book Number (ISBN) is a unique identifier for books. The European Article Number (EAN) is a broader identifier used for retail items including but beyond books, and is typically encoded in barcodes. For books, the ISBN is often embedded in the EAN.} ---against their own proprietary databases of diverse titles, and apply other automated classification rules.
Because ISBNs track specific \textit{editions}, this approach can introduce inconsistencies. 
One interviewee, for example, received a recommendation for a different edition of a book their library already owned.
% For example, one interviewee said they were recommended a different version of a book they already owned.)

\begin{figure}
    \centering
    \includegraphics[width=\columnwidth]{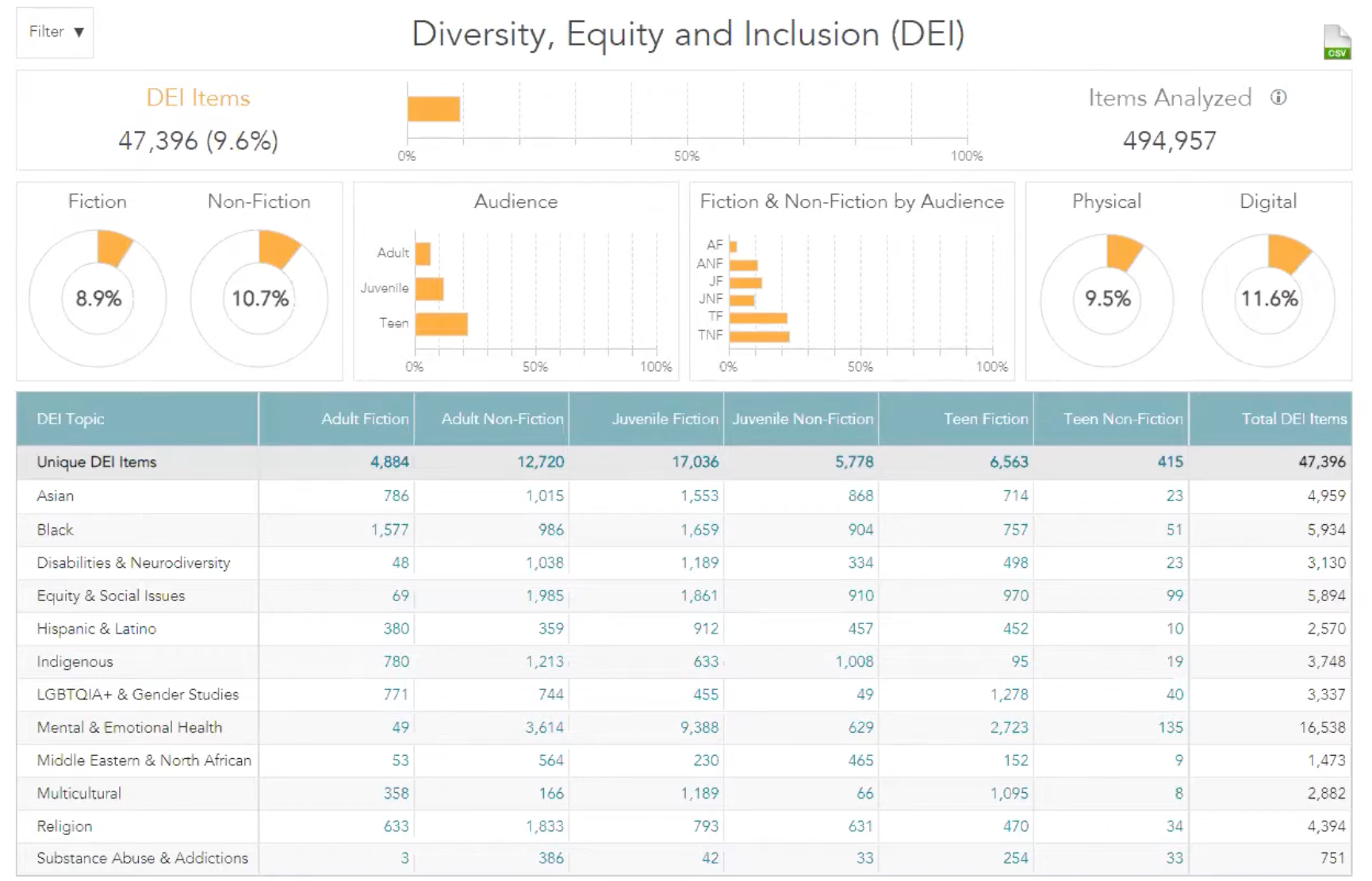}
    \caption{A dashboard featuring results from Baker \& Taylor's CollectionHQ DEI Analysis tool, showing representation of books per DEI category.}
    \label{fig:dashboard}
\end{figure}

Baker \& Taylor provides examples of how subject headings are mapped to categories: a title tagged ``Cooking, Middle Eastern; Arab American'' would be classified as ``Middle Eastern \& North African,'' and one with ``People \& Places / Canada / Indigenous''  would be classified as “Indigenous” \cite{DiversityEquityInclusiona}.
While seemingly straightforward, such classifications sometimes drew criticism from respondents, who noted the recommendation of cookbooks over more substantive works and the flattening of cultural and geographic specificity (Section~\ref{homogenize}).
Most systems allow books to appear in multiple categories. 

Table \ref{tab:vendor-tools} includes a full comparison of audit methods, price points, and scopes.
Each audit yields a quantitative report with metrics for overall diversity and specific categories, commonly presented through pie and bar charts (Figure \ref{fig:dashboard}).

Vendors often claim their audits are context-sensitive and designed to avoid reinforcing stereotypes \cite{ingramICurateInClusiveSchools2025}. 
In other words, a book featuring an inaccurate, negative, or harmful portrayal of a demographic group would not be counted as ``diverse.''
For these reasons, collectionHQ’s \textit{DEI Analysis} allows libraries to remove books from categories deemed inappropriate (and if enough do so, the book is excluded system-wide). 

Yet it remains difficult to assess how such contextual sensitivity is applied at scale. 
For example, Sharri Markson’s \textit{What Really Happened in Wuhan} (2021), a journalistic account of COVID-19’s origins, was tagged as ``Asian Interest'' in one vendor report. 
Markson is a Jewish Australian journalist, and the book’s BISAC headings include: ``HEALTH \& FITNESS / Diseases / Contagious (incl. Pandemics),'' ``POLITICAL SCIENCE / Geopolitics,'' ``SOCIAL SCIENCE / Conspiracy Theories,'' and ``TRAVEL / Asia / East / China.'' 
While the terms ``Asian'' and ``China'' appear in the BISAC headings, this book is conceptually distinct from the kinds of titles libraries often seek when addressing racial representation or community demographics. 
% This example illustrates how automated classification, driven by metadata, can obscure the interpretive decisions shaping these systems.
This example illustrates both the limitations of automated systems that rely heavily on metadata and the consequences of not making methodological choices more transparent. 

\begin{table}[ht]
\resizebox{\columnwidth}{!}{
\def\sym#1{\ifmmode^{#1}\else\(^{#1}\)\fi}

\begin{tabular}{lll}
\toprule
\textbf{Baker \& Taylor Categories} & \textbf{Ingram Categories} & \textbf{Overdrive Categories} \\
\midrule
Asian \& Pacific Islander (with subcategories)                             & Asian Interest           & Asian  \\
Black                              & Black Interest              & African American\\
Disabilities \& Neurodiversity     & Neuro and Physical Diversity & Mental \& Physical Differences \\
Hispanic \& Latino                 & Latine Interest           & Hispanic \& Latino \\
Indigenous                         & Indigenous Interest        & Indigenous \\
Sexuality \& Gender (with subcategories)      & LGBTQIA+ Interest         & LGBTQ+ \\
& & Women \\
Mental and Emotional Health        & Mental Health             & Mental Health \\
Middle Eastern \& North African    & Middle Eastern Interest & Non-US Geography  \\
& & African \\
Multicultural                      & Multicultural             & Multicultural Studies\\
Religion (with subcategories)                           & Muslim Interest          & Religion \\
Equity \& Social Issues            & Jewish Interest            & Immigration \\
Substance Abuse \& Addictions & & Urban \\
& & Class \\

& & Alternative Family \\
\bottomrule
\end{tabular}
}

\caption{Comparison of diversity audit categories by vendor. Note: OverDrive uses these categories for visualization purposes; they measure 500 BISAC subjects. Baker \& Taylor subcategories include: Asian \& Pacific Islander (Central Asian, East Asian, Pacific Islander, South Asian, Southeast Asian); Religion (Agnostic \& Atheist, Buddhist, Hindu, Muslim, Jewish); Gender \& Sexuality (Bisexual, Gay, Lesbian, Transgender).}
\label{tab:diversity-categories}
\end{table}

Additionally, while OverDrive shares a report with all 500 BISAC subjects used in their analysis, the other vendors do not disclose the exact details or criteria of their approach.
This gives the audits an aura of ``smoke and mirrors,'' according to one interviewee.
Library workers are often left to trust that audits are conducted sensitively and responsibly on tens of thousands or even millions of titles.

% Many participants appeared to lack a fundamental understanding of how these audits functioned on a basic level.
% prevented them from carefully evaluating the method and results.
% This is the proprietary information that makes their audit profitable. 
% Many participants noted that this purposeful lack of transparency  
% We also observed several participants who did not seem to understand how the audits worked on a basic level. 

%      \item \textbf{Vendor-driven diversity audits offer convenience, benchmarking, and evidence}. The diversity audits provided by \baker, 
%     \ingram, and OverDrive were generally recognized as being convenient and accurate (see Figure \ref{fig:effective}).
%     %That is, they offered librarians access to content which meets pre-determined definitions of diversity. 
%     Faced with budget and time constraints, such lists can alleviate the burden for libraries to broaden their collections. 

% \end{itemize}
\subsection{Collection Diversity Audits Are Prevalent (RQ1)}
\label{rise}

% To address \textbf{RQ1}, we assess how widely automated and manual diversity audits have been adopted by U.S. public libraries. 
We find that overall adoption of collection diversity audits seems to be increasing.
According to our survey, 67 respondents (68\%) across 24 states reported that their library had conducted a collection diversity audit. 
Of the 32 who had not, 21 respondents (70\%) said that their library will or might conduct one in the future. 
While our survey is limited in scope, these findings align with trends reported elsewhere. 
\textit{Library Journal} found that the percentage of public libraries conducting diversity audits rose from 5.5\% in 2019 to 46\% in 2022 \cite{vercellettoHowDiverseAre,wyattCollectionRebalance2022a}, suggesting that the practice has become increasingly common nationwide.

In the spending data, we see collection diversity audit purchases across at least 13 states, and adjacent purchases such as seminars, webinars, and e-courses \textit{about} diversity audits, and guidebooks \textit{for} conducting diversity audits (\ref{spending}).
Libraries spend more on other diversity-related purchases (\ref{spending})---such as staff trainings---but overall, they spend more on vendors who provide collection audits (Table \ref{tab:LibA} and \ref{tab:LibB}). 

\subsection{Vendor Audits Increase Vendor Dependence (RQ1, RQ3)}
\label{dependence}

Vendor-provided diversity audits have become a popular and often appreciated tool 
% among public libraries, in large part because they offer technical capabilities that many libraries cannot easily replicate in-house. 
but we find that they contribute to deeper forms of financial and technical dependence on vendors.
In our survey, 40 respondents (60\%) reported that their library had used a commercial audit, either exclusively or in conjunction with internal assessments. 
The most commonly used vendors include Ingram (24), Baker \& Taylor (18), OverDrive (12), Midwest Tape (2), and the non-profit Diverse BookFinder (4), with 15 respondents (37.5\%) reporting use of more than one audit. Using multiple vendor audits in a single year is not uncommon, as is confirmed by interviewees (e.g., P5, P8, P9) and public records \cite{theseattlepubliclibraryBoardTrusteesMeeting2023}.

% \begin{figure}
%     \centering
%     \includegraphics[width=\linewidth]{Figures/Baker-County-Bar-Chart.png}
%     \caption{Example report drawn from Ingram's website, showing representation per diversity category.}
%     \label{fig:bar-chart}
% \end{figure}

Twenty-four respondents (60\%) said they chose a commercial audit because of an existing relationship with the vendor. 
Audits are sometimes offered at no additional cost as part of broader service contracts, like an OverDrive or CollectionHQ subscription. 
This bundling makes vendor audits especially attractive for libraries with limited staff capacity or technical expertise to conduct in-house assessments.

\begin{figure}
    \centering
    \includegraphics[width=\linewidth]{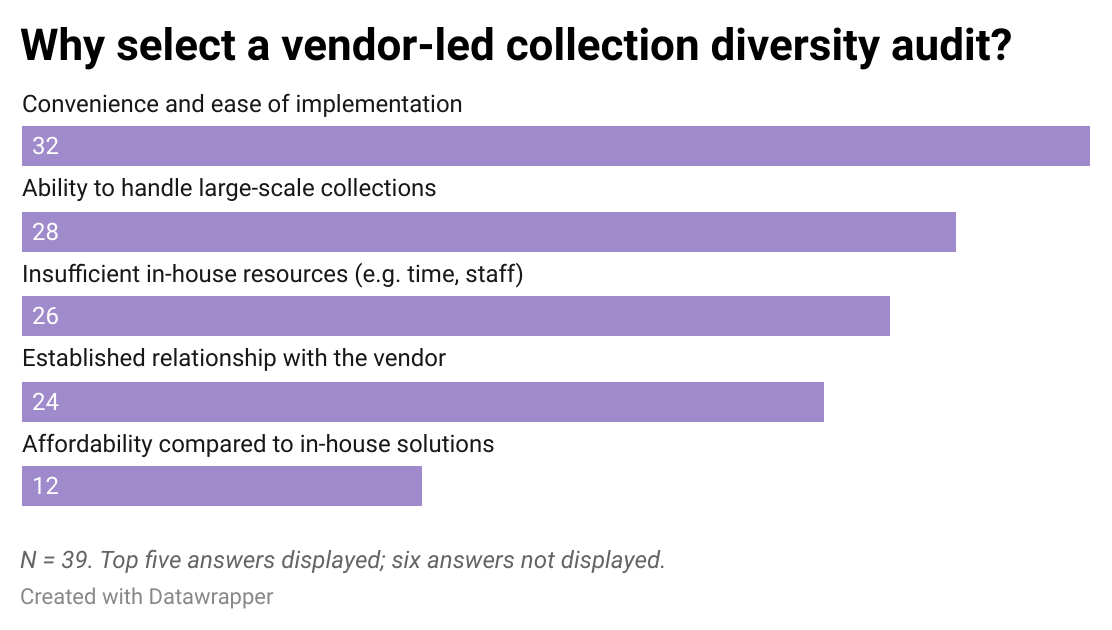}
    \caption{Reasons libraries chose to conduct a vendor-provided collection diversity audit, according to survey respondents.}
    \label{fig:effective}
\end{figure}

\begin{table}[!htb]
  \centering
  \resizebox{\columnwidth}{!}{

\begin{tabular}{lrrrrr}
\toprule
Company Name               & 2021    & 2022    & 2023    & 2024    & GovSpend Total \\
\midrule
\textbf{OverDrive}         & \textbf{\$3.1M} & \textbf{\$108K} & \textbf{\$4.2M} & \textbf{\$2.5M} & \textbf{\$9.9M} \\
Midwest Tape              & \$235K   & \$250K   & \$230K   &         & \$715K   \\
\textbf{Kanopy (OverDrive)}            & \textbf{\$110K}   & \textbf{\$120K} & \textbf{\$115K} &         & \textbf{\$345K} \\
\textbf{Baker \& Taylor}   & \textbf{\$89K}    & \textbf{\$34K}  & \textbf{\$39K}  & \textbf{\$6K}  & \textbf{\$168K} \\
ProQuest                 & \$39K    & \$41K    & \$54K    &         & \$134K   \\
\textbf{Ingram Library Services} & \textbf{\$77K}    & \textbf{\$4K}  &         &         & \textbf{\$82K}    \\
\midrule
Total GovSpend          & \$3.9M & \$839K   & \$4.9M & \$2.7M &           \\
Total Library \\ Collection Spending\textsuperscript{†} & \$8.9M & \$9.7M & \$11.7M &         &           \\
\bottomrule
\end{tabular}

% \end{adjustbox}

  }
  \caption{Spending by a large Midwestern library (Library A)  with select vendors between January 1, 2021 and August 9, 2024, according to GovSpend. Records include items over \$100. Since GovSpend records are incomplete, figures represent the minimum amount spent. \textsuperscript{†}Total library collection expenditures were obtained from state audit reports.}
  \label{tab:LibA}
\end{table}

\begin{table}[!htb]
  \centering
  \resizebox{\columnwidth}{!}{

    % \begin{adjustbox}{width=1.3\columnwidth,center}{
% \def\sym#1{\ifmmode^{#1}\else\(^{#1}\)\fi}

% \begin{tabular}{rrrrr}
% \toprule
% Company Name                      & 2022         & 2023           & 2024           & GovSpend Total    \\
% \midrule
% \textbf{OverDrive     }                & \textbf{\$103,980.87} & \textbf{\$673,184.68}   & \textbf{\$371,638.34}   & \textbf{\$1,148,803.89} \\
% Midwest Tape                      & \$81,191.19  & \$559,609.27   & \$291,663.30   & \$932,463.76   \\
% Bibliocommons                     & \$106,135.37 & \$208,844.00   & \$314,979.37   & \$629,958.74   \\
% Digital Preservation              & \$20,000.00  & \$176,166.45   & \$80,000.00    & \$276,166.45   \\
% Proquest  & \$25,555.18  & \$88,045.20    & \$16,840.51    & \$130,440.89   \\
% \bf{Ingram Library Services  }         & \textbf{\$55,118.31}  & \textbf{\$30,972.05}    & \textbf{\$17,019.49 }   & \textbf{\$103,109.85}   \\
% Bibliotheca                       & \$23,167.10  & \$8,207.30     & \$31,374.40    & \$62,748.80    \\
%                                 \midrule
% GovSpend Total                       & \$647,590.85 & \$2,746,828.25 & \$1,623,034.87 &              \\
% \bottomrule
% \end{tabular}
% }

% \end{adjustbox}

\begin{tabular}{lrrrr}
\toprule
Company Name               & 2022    & 2023    & 2024    & GovSpend Total \\
\midrule
\textbf{OverDrive}         & \textbf{\$104K} & \textbf{\$673K} & \textbf{\$372K} & \textbf{\$1.1M} \\
Midwest Tape              & \$81K   & \$560K   & \$292K   & \$932K   \\
\textbf{Ingram Library Services} & \textbf{\$55K}   & \textbf{\$31K}   & \textbf{\$17K}   & \textbf{\$103K} \\
\midrule
GovSpend Total           & \$240K   & \$1.3M   & \$681K   & \textbf{\$2.2M}      \\
\bottomrule
\end{tabular}

  }
  \caption{Spending by a large Midwestern library (Library B) with select vendors between October 11, 2022 and August 15, 2024, according to GovSpend. Records include items over \$100. Since GovSpend records are incomplete, figures represent the minimum amount spent.}
  \label{tab:LibB}
\end{table}

% \begin{table*}[!htb]
%     \parbox{\columnwidth}{
%       \centering 
%       \input{Tables/columbus}
%     \caption{Library A's spending with selected vendors between October 7, 2020 and August 9, 2024, according to GovSpend. Records include items over \$100, summarizing 3,953 purchases across 786 vendors. Library collection expenditures were obtained from state audit reports. Since GovSpend records are incomplete, the reported figures represent the minimum amount the library spent during this period.}
%     \label{tab:LibA}
%     }
%     \hfill
%     \parbox{\linewidth}{
%         \input{Tables/stl}
%         \caption{Library B's spending by selected vendors between October 11, 20222 and August 15, 2024, according to GovSpend. Records include items over \$100, summarizing 9,812 purchases from 700 vendors. Since GovSpend records are incomplete, the reported figures represent the minimum amount the library spent during this period.}
%         \label{tab:LibB}
%     }
% \end{table*}

These “free” audits are rarely without cost, however. Purchasing records from two large Midwestern library systems show that they spent hundreds of thousands—and in some cases, millions—of dollars annually on materials and services from vendors such as Ingram, OverDrive, and Baker \& Taylor (\tabref{tab:LibA}, \tabref{tab:LibB}).
These audits function as a value-added feature that reinforces existing vendor relationships and incentivizes continued purchasing. 
In some cases, libraries conduct the same audit multiple times: Library A paid Ingram an average of \$3.6K annually over three years for diversity audits.

Vendors typically recommend purchasing their own titles to fill identified diversity gaps.
The audit thus serves as both evaluator and sales driver, and sales can be substantial.
Most respondents (28\%) estimated their library spent between \$5K and \$9K per audit cycle, including new books purchased, with three (7\%) reporting totals exceeding \$30K.
After conducting three vendor audits in 2023, the Seattle Public Library acquired 1.5K new titles, totaling nearly 4.7K print and digital copies  \cite{theseattlepubliclibraryBoardTrusteesMeeting2023}. 

Fifty respondents (50\%) expressed that they were \textit{somewhat} or \textit{very} concerned that vendors might have a conflict of interest when auditing and recommending their own titles for purchase.
% When asked about potential conflicts of interest with vendor-driven audits, especially regarding the recommendation of books to fill diversity gaps, many respondents did not see major issues. 
Others were less concerned, though sometimes due to incomplete information. 
As one participant noted, ``The vendor we used, collectionHQ, does not sell books. We use them to run reports that tell us stats about the items in our collection. So they have no reason to have a conflict of interest.'' 
% In reality, however, CollectionHQ is owned by Baker \& Taylor, and so they do, in a sense, sell books. 
In fact, collectionHQ is owned by Baker \& Taylor—one of the largest book distributors in the U.S.
This response points to a gap in awareness about the increasingly complex and opaque relationships among vendors, especially in an era of industry consolidation. 

\subsection{Vendor Audits Homogenize Diversity---And Potentially Collections (RQ2, RQ3)} \label{homogenize}
\label{homog}

% Even before collection diversity audits, there have been concerns that large library vendors homogenize collections \citep{wallace1997outsourcing,hoffert2007s}. 
% Librarians in Hawaii mounted a grass-roots campaign against \baker{} \citep{knuth1998revolt}, which 
% ultimately resulted in a series of legal actions between \baker{} and various government entities \citep{farmanfarmaian1997hawaii}.

% To better understand how vendor-driven diversity audits actually function (RQ2) and how library workers perceive their usefulness and nuance (RQ3), we analyzed participants’ experiences with the categorical frameworks used by vendors. 

Many librarians expressed frustration with the overly broad and opaque nature of the diversity categories in automated diversity audits (Table \ref{tab:diversity-categories}). 
These standardized labels often failed to capture local demographic nuances. 
Thirty-two respondents (41\%) said they were \textit{not too} or \textit{not at all} effective at being responsive to community needs.
Some librarians perceived them as reductive or even counterproductive.
For example, P14, a librarian serving a Tribal Nation, shared dissatisfaction with the umbrella term ``Indigenous'':

\begin{quote}
[Vendors] assume that a book about a Cherokee experience has any contact with the Mandan people in North Dakota. So that's where the diversity labeling gets into a sticky mess. The granular level is the better level because you're much more localized. But that's not how the big companies do it. 
\end{quote}

The same critique was also raised in the survey responses: ``To bundle `Indigenous' materials as a single concept is akin to bundling `European' as a concept: inadequate and superficial.'' 
These frustrations echo scholarly critiques about the flattening of Indigenous identities and knowledge within LCSH \cite{duarteImaginingCreatingSpaces2015}.
Respondents also flagged the use of the category ``Black'' as overly broad—failing, for example, to distinguish between African American and African or African diasporic literatures (one respondent noted the audit neglected ``a HUGE collection of African works of literature'').

Interviewees wished for more customization related to specific local communities.
For example, P7 conducted an in-house diversity audit to assess how well their collection was supporting their community's large and growing Bengali population, but vendor-driven audits did not allow for this level of granularity in their diversity categories.\footnote{Some audits, like Baker \& Taylor's, have recently introduced more subcategories--such as ``South Asian,'' which includes Bengali---but the granularity is still limited overall.}

% Librarians also raised questions about other broad categories, some of which seemed like a distraction from more meaningful diversity work.
% For example, P2 called ``Mental Health'' was ``broad, broad, broad'' and ``Multicultural'' a ``catchall.'' 
% P8 shared frustration about co-workers who tended to focus on easier categories to select over more challenging ones, such as mental health over race and sexuality: ``They'll latch onto that, and they won't deal with the tough topics.'' 

% Expressing confusion over certain categories, P2 states `` `Mental Health' is very broad, broad, broad. And then `Multicultural?' That's a very much, a catch-all. I don't know. And, you know, that's their category. What are you basing that on? You know, that's not as helpful as some other categories.'' 

Other respondents took issue with the actual books that were recommend in these broad categories, reporting that they seemed to be superficially or even offensively related. 
One librarian said that they were looking for Indigenous and Middle Eastern nonfiction but ended up getting recommended mostly cookbooks. 
Another asked why ``one of the big vendors put[s] books about basketball only into their `Black interest' category and not other categories?''

\subsection{Vendor Audits Lack Transparency---and Sometimes Trust (RQ2, RQ3)}
\label{transparency}

% To further address RQ2 and RQ3, we examine how vendor audits operate and how library workers perceive their accuracy, transparency, and reliability.
We find that most audits lack methodological transparency, which undermines trust in results for some library workers.
Interviewees consistently exhibited confusion, skepticism, or lack of knowledge about how audit results were generated. 
% Many noted that vendors do not disclose the specific criteria or data used to classify books into diversity categories.
% —particularly when those classifications rely on subject headings or proprietary datasets. 
% As a result, library workers often felt unable to evaluate or verify the audit findings.
While most survey respondents thought vendor audits were \textit{about the same}, \textit{more}, or \textit{much more} trustworthy than those conducted by internal  staff, thirty-six (40\%) answered that they were \textit{less} or \textit{much less} trustworthy. 
% When asked how trustworthy vendor audits were compared to those conducted by internal library staff or independent third-party organizations, most respondents thought they were about the same or more trustworthy. 
% However, a notable 34\% answered that they were less trustworthy (\figref{fig:trust}). 

These concerns were compounded by broader misgivings about metadata quality. 
Several librarians observed that subject headings—the core of many vendor audit systems—are often incomplete, inconsistently applied, or outdated. 
P7, for example, raised questions about whether metadata could accurately reflect the diversity of a book's content:

\begin{quote}
I've worked with metadata. I know how bad it can be... a lot of the MARC  [Machine-Readable Cataloging] records for books in ILS [Integrated Library System] systems or in book vendors, those subject headings are assigned by publishers who aren't necessarily librarians. And then now we're running a diversity audit on that? So what are we missing?
\end{quote}

% P7 went on to describe an early reader series featuring a young Indigenous girl. Because the series centered on themes of friendship and daily life, the books likely would not be flagged as “Indigenous” based on subject headings alone—raising concerns about meaningful inclusion being missed by automated tools.

These metadata issues were frequently linked to a broader decline in vendor service quality following the COVID-19 pandemic. Interviewees reported that supply chain disruptions, staff turnover, and rising costs all contributed to a reduction in cataloging quality \cite{IngramCOVIDSupplyChain}. 
% In one case, a major vendor also experienced a ransomware attack that further delayed services and disrupted workflows \cite{IngramCOVIDSupplyChain,BreedingBaker&TaylorOutage}.
As vendor audits increasingly depend on automated classification and bibliographic metadata, the deterioration in catalog record quality raises serious questions about the reliability of these tools.

\subsection{Vendor Audits Offer Convenience, Benchmarking, and Evidence (RQ3)}
\label{convenience}
% Finally, we explore the perceived benefits of vendor-driven diversity audits as reported by library workers. 
Despite the critiques around transparency and category design, many interviewees emphasized the convenience, time savings, and institutional utility of these tools. 
Seventy respondents (90\%) said they were \textit{somewhat} or \textit{very} effective at delivering quick, scalable results.
Sixty-four (82\%) said they were effective at producing evidence to communicate with stakeholders.
Nine interviewees similarly mentioned time saved and convenience as a major draw. 
``There's no way we would've had time to manually go through even a portion of our collection,'' P9 shared.
Librarians reported that vendor-driven audits not only streamline the evaluation process, but also provide evidence that can guide collection development, demonstrate commitments to administrators, and support broader DEI initiatives.
% ---such as outreach strategies, event programming, updating internal policies and trainings, and evaluating book displays.

Participants also found that comparative benchmarks---showing their results compared to ``average'' public library results---were useful not only for highlighting areas of improvement but for demonstrating to critics that their collections were not abnormally imbalanced.
One motivation of their diversity audit, P4 said, was to prove ``we're not that crazy. We're not that different than other libraries. It's not like we're... just offering wildly crazy books that no one else is offering.''
This sentiment has also been expressed by other library workers.
``The data shows the pushback is not founded in reality,'' one New York librarian told journalists after a diversity audit. 
``Our collections aren’t what they think they are''  \cite{escuderoHowAlbanyAdding2023}.
% Most early diversity audits were conducted manually or at a small scale, often with librarians physically pulling items from shelves, examining book covers, reading descriptions, and more. 

% benchmark While desiring more nuance and customization, interviewees also appreciated that vendor-driven audits provide some benchmarking for areas of focus. 
% Interviewees find it useful to compare their results to local demographics and peer institutions, especially when run periodically to track changes made as a result of initial audit findings. 
% A cyclical approach to diversity audits proves difficult for libraries, which lack the resources to conduct one in-house audit, let alone recurring studies, while also likely lacking funds to pay for vendor-driven audits regularly.

\section{DISCUSSION}

Our findings reveal that for many library workers, automated collection diversity audits provided by commercial vendors offer a useful and time-saving tool for addressing a real and urgent problem: the lack of diversity in library collections. 
As we found in Section \ref{convenience}, library workers reported that these tools deliver quick results, generate stakeholder-facing reports, and provide curated lists for acquisition—filling a gap for many libraries that lack staff capacity to conduct in-house assessments.
The standardization of diversity categories also facilitates comparisons across institutions and provides evidence that libraries can present to stakeholders and the broader community---including advocacy groups bent on limiting access to diverse books.

We affirm that the lack of various kinds of diversity in library collections is an urgent problem. We believe that automated tools, when thoughtfully designed and deployed, may serve as effective mechanisms for addressing large collections that cannot be manually reviewed title-by-title. 
% In contexts where “perfect is the enemy of good,” even imperfect tools may be better than no tools at all. 
As political attacks on diversity and libraries escalate, it is important to recognize that what may seem inadequate from a distance may nonetheless offer vital support to library workers on the ground \cite{schermeleCulturalPowerStruggle2022, escuderoHowAlbanyAdding2023}.

However, our study surfaces several major concerns in existing vendor collection diversity audits, in the broader framing of diversity they promote, and in public libraries' deepening financial and technical dependence on conglomerate vendors. 
When considered in light of related work on algorithmic bias, fairness, critical race theory, and library privatization, additional concerns emerge.

First, as detailed in Sections \ref{categories} and \ref{homog}, library workers expressed concerns about the homogeneity of diversity categories and the disconnect between standardized labels and the specific needs of their local communities. 
One respondent compared the bundling of “Indigenous” materials to labeling all “European” books under one heading, pointing to the loss of nuance, history, and identity in such taxonomies.
Thirty-two respondents (41\%) said they were \textit{not too }or \textit{not at all} effective at being responsive to community needs.
% Other examples—like cookbooks being flagged as Indigenous nonfiction or basketball books categorized solely as “Black Interest”—highlight how metadata-driven classification can misfire in ways that are not merely imprecise, but actively distortive.

% that efficiency and profitability too often come at the expense of cultural specificity and community representation. 
% Library workers have long warned that large vendors tend to homogenize collections, often prioritizing efficiency and profitability over local relevance. In the late 1990s, for example, librarians in Hawai‘i mounted a grassroots campaign against \baker{}, arguing that its centralized collection development processes failed to meet the needs of Native Hawaiian communities \citep{wallace1997outsourcing, hoffert2007s, knuth1998revolt, farmanfarmaian1997hawaii}. As one librarian, Deborah Gutermuth, put it: “The concept as a whole is flawed. It’s like putting Safeway in charge of the school lunch program. The kids will be fed what makes a profit.” At the time, state librarian Bart Kane defended the decision to outsource, citing severe budget cuts and warning that without such a contract, Hawai‘i’s library system could have faced mass closures and layoffs. 
% Today’s critiques mirror these earlier tensions. 

Second, as discussed in Section~\ref{dependence}, vendor audits reinforce libraries’ financial and infrastructural dependence on commercial providers. 
Public libraries fundamentally rely on commercial vendors for a wide range of services, but these audits may deepen vendor lock-in and disincentivize libraries from developing their own automated or manual assessment tools, or from investing in human time and labor. 
% Audits are typically bundled into costly contracts for books, digital materials, and collection management platforms. 
As libraries cede more authority over content, analytics, and infrastructure to commercial providers, they may lose the capacity—and bargaining power—to evaluate and shape their own collections independently.
% Our procurement analysis (Tables \ref{tab:LibA} and \ref{tab:LibB}) shows that large library systems spend hundreds of thousands—or even millions—annually on services from vendors like Ingram, OverDrive, and Baker & Taylor. 

% while also directing libraries to purchase the very books the vendor classifies and supplies.

Third, market competition discourages vendors from fully disclosing how their tools work, limiting transparency and contributing to confusion or mistrust among library workers (Section~\ref{transparency}). 
% Many reported that vendors do not disclose the classification logic behind their category assignments, and several noted inconsistencies or mismatches between what vendors flagged as “diverse” and their own local knowledge of the collection. 
These concerns were compounded by broader skepticism about the quality and provenance of bibliographic metadata. 
In total, 36 (40\%) respondents said they trusted vendor-conducted audits \textit{less} or \textit{much less} than audits conducted by internal staff.
% As one respondent observed, publishers—not librarians—often assign BISAC codes, which vendors then use to drive audit logic. In such a system, audit accuracy depends on opaque, inconsistently applied metadata pipelines that libraries cannot control or verify.

Notably, many of these critiques are not new. 
For example, in the late 1990s, librarians in Hawai‘i mounted a grassroots campaign against Baker \& Taylor after the state librarian signed a large outsourcing contract for collection development. 
Opponents argued that the vendor’s privatized, centralized approach failed to meet the needs of Native Hawaiian communities \citep{wallace1997outsourcing, knuth1998revolt}. 
As one librarian warned, ``It’s like putting Safeway in charge of the school lunch program. The kids will be fed what makes a profit'' \cite{wallace1997outsourcing, hoffert2007s, knuth1998revolt, farmanfarmaian1997hawaii}.
Today’s critiques mirror these earlier tensions.
Automated diversity audits may ultimately extend rather than resolve the longstanding problem of vendor-driven and profit-driven library collection standardization.

These findings also resonate with broader critiques about the operationalization of race and the cooption of diversity. As \citet{hannaCriticalRaceMethodology2020} argue, “treating race as an attribute, rather than a structural, institutional, and relational phenomenon… minimize[s] the structural aspects” of inequality. In the context of libraries, these structural forces include systemic discrimination in the publishing industry, the commercial consolidation of the book industry, and the chronic underfunding of public institutions. 
In a similar vein, \citet{melamedRepresentDestroyRationalizing2011} argues that frameworks of diversity were coopted and commodified after World War II, and that they mask and enable ongoing racial exploitation; she even specifically highlights diverse literature as a central instrument in this process. 
% \citet{melamedRepresentDestroyRationalizing2011}.
% She even specifically highlights the role of literature as a key instrument in this dynamic.
% This dynamic echoes the critique advanced by \citet{melamedRepresentDestroyRationalizing2011}, who shows how diversity has been coopted and commodified after World War II, and how it masked, and even enabled, ongoing racialized exploitation. 
% She even specifically pinpoints book as a central instrument in this project.
These perspectives suggest that diversity audits risk turning numerical targets into an end in themselves---pursued without regard for the quality, relevance, or political significance of the selected materials, and without prompting meaningful changes to collection development practices or engagement with community knowledge

Moreover, the widespread adoption of automated audits risks hollowing out the critical, context-specific labor that collection development traditionally entails. 
% By outsourcing essential evaluative and ethical decisions to vendors, libraries risk becoming passive nodes in privatized information pipelines rather than active stewards of their collections. 
% This shift reflects and accelerates broader trends in information infrastructure \citet{lamdanDataCartelsCompanies2022}.
Without careful intervention, libraries may find themselves increasingly dependent on opaque external systems, alienated from the intellectual and social responsibilities at the heart of their mission.

% Finally, as several interviewees noted,vendor audits can only reflect the publishing pipelines they’re built upon. If vendors primarily source from mainstream, U.S.-based, English-language publishers, their diversity assessments will always be constrained by those pipelines. As one librarian put it, “We're just buying what we would have bought anyway, and just trying to be a little more thoughtful.” This raises a final imperative: that libraries not only scrutinize vendor tools, but also advocate for more equitable publishing ecosystems.
% Addressing these systemic issues requires collective advocacy beyond libraries themselves, targeting the structures that shape what books get published, marketed, and made available in the first place.

\subsection{Recommendations} 
Automated collection diversity audits represent just one among many data-driven, algorithmic, and AI tools now being introduced to libraries. 
% There are also AI-powered recommendation systems, circulation prediction algorithms, search and discovery tools, chatbot assistants, and many other services likely being developed.
Based on our findings, we present the following recommendations for automated diversity audits, emphasizing that many of these principles extend to libraries’ adoption of automated and AI systems more broadly.

\begin{itemize} \item \textbf{Enhance Transparency in Automated Vendor Tools}: To improve trust and understanding, vendors should fully disclose classification criteria, data sources, and results. As shown in Section \ref{transparency}, library workers often do not know exactly how audit classifications are made. More than a third of respondents also found them less trustworthy than those performed by library staff. As AI systems become more embedded in library workflows, opaque infrastructures will only deepen unless transparency becomes a collective demand.
% Libraries may benefit from academic partners and professional organizations to interpret methodologies.

\item \textbf{Promote Flexibility and Contextual Awareness}: Baker \& Taylor's tool currently offers the most customization, but vendor tools should allow more customization and crowdsourcing. Section \ref{homogenize} shows that libraries want tools that reflect local demographic complexity. 

\item \textbf{Invest in Staffing and Manual Review Processes}: Libraries should secure funding and create positions for manual assessment, framing this labor as core—not supplemental—to ethical data practices. Throughout our interviews, librarians consistently emphasized the importance of human judgment and review. 

\item \textbf{Support Open Alternatives and Build Internal Data Capacity}: As discussed in Sections \ref{dependence}, vendor audits further cement library relationships with vendors. Reducing dependence on proprietary systems requires investment in free, open-source tools and broader data capacity. For example, the Institute of Museum and Library Services (IMLS) has funded tools like Diverse BookFinder, which supports audits of diversity in picture books. Graduate programs in library science might also incorporate more training in data literacy and computational methods.

\item \textbf{Organize Collectively to Influence Vendor Practices}: 
Because most libraries lack bargaining power on their own, regional consortia, professional associations, and advocacy organizations—such as the \href{https://www.ala.org/}{American Library Association (ALA)}, \href{https://sparcopen.org/}{SPARC}, and \href{https://libraryfutures.net/}{Library Futures}—should collaborate to demand greater transparency, flexibility, and accountability in automated vendor tools. 
These groups have already laid important groundwork by promoting open infrastructures, equitable access, and public-interest technology policy. 

\end{itemize}

\subsection{The Future of Collection Diversity Audits---And Public Libraries}
% We also worry about a political climate in which libraries are increasingly under-resourced and attacked, and thus ever more reliant on automation and third-party vendor services.
The future of diversity audits remains uncertain as a wave of reactionary anti-DEI and anti-library action has overtaken institutions across the U.S. 
In early 2025, the Trump administration made drastic cuts to the Institute of Museum and Library Services (IMLS)—a key source of federal funding for libraries, programs, and innovation nationwide—as well as to several state libraries \citep{EO4, IMLS_cuts}. 
In May, Trump fired Carla Hayden, the first Black Librarian of Congress, and his administration cited her ``pursuit of DEI'' and supposed placement of ``inappropriate books in the library for children'' as justification \cite{nguyenTrumpFiresLibrary2025,limbongLibrarianCongressFiring2025}.

% issued an executive order to drastically reduce funding to the IMLS \citep{EO4}. The president also appointed Keith E. Sonderling as acting director of the IMLS, who recently placed all 70 employees on administrative leave, halting any future grants, and likely terminating existing ones \citep{IMLS_cuts}. 
% IMLS not only funds innovation---like Diverse BookFinder---but also libraries around the country.
Even before Trump's second election, conservative state governments had forced libraries to withdraw their membership from ALA \citep{StatesLeaveALA}, banned DEI initiatives \citep{FloridaDEIban}, and threatened librarians with  \$5,000 or jail time for ``distribution of harmful materials to minors'' \citep{GeorgiaLaw}. 
This is among 128 other bills levied against libraries in 2024, with almost as many levied five months into 2025 \citep{2024BillTracking, 2025BillTracking}. 
Additionally, Moms for Liberty, a nonprofit with over 130,000 volunteer members in 49 states, continues to organize book challenges and bans \citep{AbooutMomsforLiberty}. 
Lastly, after supporting DEI initiatives in the wake of 2020, many mainstream publishers have backpedaled, and several prominent Black women in the industry have been fired, causing outcry and calling into question the availability of diverse books moving forward \citep{Sinykin_So_2024, alterLotUsAre2024}. 

Whether public libraries will be able to continue conducting collection diversity audits, at least under this name, is unclear.
We speculate that budget cuts may heighten libraries' need to outsource labor to automated tools, and that anti-DEI legislation will hamper efforts to improve collection diversity.

\section{CONCLUSION}

We explore the benefits and limitations of automated library collection diversity audits by drawing on a survey, interviews, purchasing records, and an analysis of vendor documentation.
% We believe that collection diversity audits are an important case study for considering how, if at all, library work and other cultural work can be automated equitably and transparently.
We find that these audits arise from a real need: many library systems are so large and resource-constrained that some form of computational assistance is necessary to assess the inclusivity of their collections.
Vendor-led audits offer a convenient but homogenizing solution—one that risks flattening the complexities of identity into standardized categories while deepening libraries’ infrastructural dependence on opaque, commercial systems.
As algorithmic tools become more embedded in public knowledge institutions, we advocate for increased transparency and flexibility, greater investment in open-source tools and human expertise, and collective organizing.

This work is especially urgent in a political moment marked by renewed attacks on libraries, education, and marginalized communities. 
In the context of rising authoritarianism and right-wing censorship efforts, we need genuine engagement with and protection of diversity. 
We must support libraries in resisting tech cooption and in ensuring automation serves the public good.

\begin{acks} 
We thank the library workers who took the time to share their experiences and perspectives with us.
We are grateful for feedback from our anonymous reviewers and from Lorcan Dempsey, Anna Preus, and others. 
This paper benefited from conversations with UCLA’s Critical Data Lab in Fall 2024.
This research was partly supported by a Strategic Research Fund award from the Information School at the University of Washington. 
\end{acks}

\bibliographystyle{ACM-Reference-Format}
\bibliography{references}


\begin{thebibliography}{129}

%%% the Web address.

\ifx \showCODEN    \undefined \def \showCODEN     #1{\unskip}     \fi
\ifx \showDOI      \undefined \def \showDOI       #1{#1}\fi
\ifx \showISBNx    \undefined \def \showISBNx     #1{\unskip}     \fi
\ifx \showISBNxiii \undefined \def \showISBNxiii  #1{\unskip}     \fi
\ifx \showISSN     \undefined \def \showISSN      #1{\unskip}     \fi
\ifx \showLCCN     \undefined \def \showLCCN      #1{\unskip}     \fi
\ifx \shownote     \undefined \def \shownote      #1{#1}          \fi
\ifx \showarticletitle \undefined \def \showarticletitle #1{#1}   \fi
\ifx \showURL      \undefined \def \showURL       {\relax}        \fi
% invisible to TeX
\providecommand\bibfield[2]{#2}
\providecommand\bibinfo[2]{#2}
\providecommand\natexlab[1]{#1}
\providecommand\showeprint[2][]{arXiv:#2}

\bibitem[Ing(2021)]%
        {IngramCOVIDSupplyChain}
 \bibinfo{year}{2021}\natexlab{}.
\newblock \showarticletitle{Ingram Warns of 4th-Quarter Supply Chain Disruptions}.
\newblock \bibinfo{journal}{\emph{Shelf Awareness Newsletter}} (\bibinfo{date}{24 Aug.} \bibinfo{year}{2021}).
\newblock
\urldef\tempurl%
\url{https://www.shelf-awareness.com/issue.html?issue=4056#m53553}
\showURL{%
\tempurl}


\bibitem[PLS(2022)]%
        {PLS2022}
 \bibinfo{year}{2022}\natexlab{}.
\newblock \bibinfo{booktitle}{\emph{Public Libraries Survey}}.
\newblock \bibinfo{type}{{T}echnical {R}eport}. \bibinfo{institution}{Institute of Museum and Library Services}.
\newblock
\urldef\tempurl%
\url{https://www.imls.gov/research-evaluation/data-collection/public-libraries-survey}
\showURL{%
\tempurl}


\bibitem[OCL(2023)]%
        {OCLCIntroducesAIgenerated2023}
OCLC \bibinfo{year}{2023}\natexlab{}.
\newblock \bibinfo{booktitle}{\emph{{{OCLC}} Introduces {{AI-generated}} Book Recommendations in {{WorldCat}}.Org and {{WorldCat Find}} Beta}}.
\newblock OCLC.
\newblock
\urldef\tempurl%
\url{https://www.oclc.org/en/news/releases/2023/20230621-ai-book-recs-worldcatorg.html}
\showURL{%
\tempurl}


\bibitem[ALA(2024)]%
        {ALA2024BannedBookData}
 \bibinfo{year}{2024}\natexlab{}.
\newblock \showarticletitle{2024 Preliminary Book Ban Data}.
\newblock \bibinfo{journal}{\emph{American Library Association's Office of Intellectual Freedom}} (\bibinfo{date}{March} \bibinfo{year}{2024}).
\newblock
\urldef\tempurl%
\url{https://www.ala.org/bbooks/book-ban-data}
\showURL{%
\tempurl}


\bibitem[Lib(2024)]%
        {LibraryProfessionalsFactsa}
Department for Professional Employees, AFL-CIO \bibinfo{year}{2024}\natexlab{}.
\newblock \bibinfo{booktitle}{\emph{Library {{Professionals}}: {{Facts}}, {{Figures}}, and {{Union Membership}}}}.
\newblock Department for Professional Employees, AFL-CIO.
\newblock
\urldef\tempurl%
\url{https://web.archive.org/web/20250429180949/https://www.dpeaflcio.org/factsheets/library-professionals-facts-and-figures}
\showURL{%
\tempurl}


\bibitem[Div(2025)]%
        {DiversityEquityInclusiona}
Baker and Taylor \bibinfo{year}{2025}\natexlab{}.
\newblock \bibinfo{booktitle}{\emph{Diversity, {{Equity}} \& {{Inclusion Resources}}}}.
\newblock Baker and Taylor.
\newblock
\urldef\tempurl%
\url{https://web.archive.org/web/20250429145154/https://www.baker-taylor.com/library-solutions/books-av-content/dei-sel}
\showURL{%
\tempurl}


\bibitem[Get(2025)]%
        {GettingStartedPrimo2024}
Ex Libris Knowledge Center \bibinfo{year}{2025}\natexlab{}.
\newblock \bibinfo{booktitle}{\emph{Getting {{Started}} with {{Primo Research Assistant}}}}.
\newblock Ex Libris Knowledge Center.
\newblock
\urldef\tempurl%
\url{https://web.archive.org/web/20250429180155/https://knowledge.exlibrisgroup.com/Primo/Product_Documentation/020Primo_VE/Primo_VE_(English)/015_Getting_Started_with_Primo_Research_Assistant}
\showURL{%
\tempurl}


\bibitem[Abdu et~al\mbox{.}(2023)]%
        {abduEmpiricalAnalysisRacial2023a}
\bibfield{author}{\bibinfo{person}{Amina~A. Abdu}, \bibinfo{person}{Irene~V. Pasquetto}, {and} \bibinfo{person}{Abigail~Z. Jacobs}.} \bibinfo{year}{2023}\natexlab{}.
\newblock \showarticletitle{An {{Empirical Analysis}} of {{Racial Categories}} in the {{Algorithmic Fairness Literature}}}. In \bibinfo{booktitle}{\emph{Proceedings of the 2023 {{ACM Conference}} on {{Fairness}}, {{Accountability}}, and {{Transparency}}}} \emph{(\bibinfo{series}{{{FAccT}} '23})}. \bibinfo{publisher}{Association for Computing Machinery}, \bibinfo{address}{New York, NY, USA}, \bibinfo{pages}{1324--1333}.
\newblock
\showISBNx{9798400701924}
\urldef\tempurl%
\url{https://doi.org/10.1145/3593013.3594083}
\showDOI{\tempurl}


\bibitem[Adams and Chiwaya(2024)]%
        {adamsMapSeeWhich2024}
\bibfield{author}{\bibinfo{person}{Char Adams} {and} \bibinfo{person}{Nigel Chiwaya}.} \bibinfo{year}{2024}\natexlab{}.
\newblock \showarticletitle{Map: {{See}} Which States Have Introduced or Passed Anti-{{DEI}} Bills}.
\newblock \bibinfo{journal}{\emph{NBC News}} (\bibinfo{date}{March} \bibinfo{year}{2024}).
\newblock
\urldef\tempurl%
\url{https://www.nbcnews.com/data-graphics/anti-dei-bills-states-republican-lawmakers-map-rcna140756}
\showURL{%
\tempurl}


\bibitem[Ahmed(2012)]%
        {ahmedBeingIncludedRacism2012}
\bibfield{author}{\bibinfo{person}{Sara Ahmed}.} \bibinfo{year}{2012}\natexlab{}.
\newblock \bibinfo{booktitle}{\emph{On Being Included: Racism and Diversity in Institutional Life}}.
\newblock \bibinfo{publisher}{Duke University Press}, \bibinfo{address}{Durham}.
\newblock
\showISBNx{978-0-8223-9532-4}


\bibitem[Alfonseca(2024)]%
        {alfonsecaMapImpactAntiDEI2024}
\bibfield{author}{\bibinfo{person}{Kira Alfonseca}.} \bibinfo{year}{2024}\natexlab{}.
\newblock \showarticletitle{Map: {{The}} Impact of Anti-{{DEI}} Legislation}.
\newblock \bibinfo{journal}{\emph{ABC News}} (\bibinfo{date}{April} \bibinfo{year}{2024}).
\newblock
\urldef\tempurl%
\url{https://abcnews.go.com/US/map-impact-anti-dei-legislation/story?id=108795967}
\showURL{%
\tempurl}


\bibitem[Alter and Harris(2024a)]%
        {alterLotUsAre2024}
\bibfield{author}{\bibinfo{person}{Alexandra Alter} {and} \bibinfo{person}{Elizabeth Harris}.} \bibinfo{year}{2024}\natexlab{a}.
\newblock \showarticletitle{`{{A Lot}} of {{Us Are Gone}}': {{How}} the {{Push}} to {{Diversify Publishing Fell Short}}}.
\newblock \bibinfo{journal}{\emph{The New York Times}} (\bibinfo{date}{Aug.} \bibinfo{year}{2024}).
\newblock
\showISSN{0362-4331}


\bibitem[Alter and Harris(2024b)]%
        {alterPublishingPledgedDiversify2024a}
\bibfield{author}{\bibinfo{person}{Alexandra Alter} {and} \bibinfo{person}{Elizabeth~A. Harris}.} \bibinfo{year}{2024}\natexlab{b}.
\newblock \showarticletitle{Publishing {{Pledged}} to {{Diversify}}. {{Change Has Been Slow}}.}
\newblock  (\bibinfo{year}{2024}).
\newblock
\showISSN{0362-4331}
\urldef\tempurl%
\url{https://www.nytimes.com/2024/02/28/books/publishing-books-poc-dei.html}
\showURL{%
\tempurl}
\newblock
\shownote{Accessed: 2025-01-22}.


\bibitem[America(2024)]%
        {penamericaBannedUSAShelves2024}
\bibfield{author}{\bibinfo{person}{PEN America}.} \bibinfo{year}{2024}\natexlab{}.
\newblock \bibinfo{title}{Banned in the {{USA}}: {{Beyond}} the {{Shelves}}}.
\newblock \bibinfo{howpublished}{https://pen.org/report/beyond-the-shelves/}.
\newblock


\bibitem[Atterbury(2023)]%
        {StatesLeaveALA}
\bibfield{author}{\bibinfo{person}{Andrew Atterbury}.} \bibinfo{year}{2023}\natexlab{}.
\newblock \showarticletitle{Florida joins conservative states severing ties with national library group}.
\newblock  (\bibinfo{date}{31 Oct.} \bibinfo{year}{2023}).
\newblock
\urldef\tempurl%
\url{https://www.politico.com/news/2023/10/31/florida-conservative-national-library-ala-00124516}
\showURL{%
\tempurl}


\bibitem[Avram(1975)]%
        {avramMARCItsHistory1975}
\bibfield{author}{\bibinfo{person}{Henriette~D. Avram}.} \bibinfo{year}{1975}\natexlab{}.
\newblock \bibinfo{booktitle}{\emph{{{MARC}}; Its {{History}} and {{Implications}}}}.
\newblock \bibinfo{type}{{T}echnical {R}eport}. \bibinfo{institution}{Superintendent of Documents, U}.
\newblock


\bibitem[{Baker \& Taylor}(2022)]%
        {baker&taylorIntroducingBakerTaylors2022}
\bibfield{author}{\bibinfo{person}{{Baker \& Taylor}}.} \bibinfo{year}{2022}\natexlab{}.
\newblock \bibinfo{title}{Introducing {{Baker}} \& {{Taylor}}'s {{Diversity Analysis Tool}}}.
\newblock
\newblock
\urldef\tempurl%
\url{https://www.youtube.com/watch?v=ldybXE2dozM}
\showURL{%
\tempurl}


\bibitem[{Baker {and} Taylor}(2025)]%
        {bakerandtaylorBakerTaylorDiversity2025}
\bibfield{author}{\bibinfo{person}{{Baker {and} Taylor}}.} \bibinfo{year}{2025}\natexlab{}.
\newblock \bibinfo{title}{Baker \& {{Taylor}} {\textbar} {{Diversity}}, {{Equity}} \& {{Inclusive}} {\textbar} {{Social}} and {{Emotional Learning}}}.
\newblock \bibinfo{howpublished}{https://www.baker-taylor.com/library-solutions/books-av-content/dei-sel}.
\newblock


\bibitem[Baron and Broadley(2019)]%
        {baronChangeSubject2019}
\bibfield{author}{\bibinfo{person}{Jill Baron} {and} \bibinfo{person}{Sawyer Broadley}.} \bibinfo{year}{2019}\natexlab{}.
\newblock \bibinfo{title}{Change the {{Subject}}}.
\newblock
\newblock
\urldef\tempurl%
\url{https://collections.dartmouth.edu/archive/object/change-subject/change-subject-film}
\showURL{%
\tempurl}


\bibitem[Benjamin(2019)]%
        {benjaminRaceTechnologyAbolitionist2019}
\bibfield{author}{\bibinfo{person}{Ruha Benjamin}.} \bibinfo{year}{2019}\natexlab{}.
\newblock \bibinfo{booktitle}{\emph{Race {{After Technology}}: {{Abolitionist Tools}} for the {{New Jim Code}}}}.
\newblock \bibinfo{publisher}{Polity}, \bibinfo{address}{Cambridge, UK Medford, MA}.
\newblock
\showISBNx{978-1-5095-2640-6}


\bibitem[Berman(2013)]%
        {bermanPrejudicesAntipathiesTract2013}
\bibfield{editor}{\bibinfo{person}{Sanford Berman}} (Ed.). \bibinfo{year}{2013}\natexlab{}.
\newblock \bibinfo{booktitle}{\emph{Prejudices and {{Antipathies}}: {{A Tract}} on the {{LC Subject Heads Concerning People}}} (\bibinfo{edition}{reprint edition} ed.)}.
\newblock \bibinfo{publisher}{McFarland \& Company}, \bibinfo{address}{Jefferson, N.C.}
\newblock
\showISBNx{978-0-7864-9352-4}


\bibitem[Billey et~al\mbox{.}(2024)]%
        {billeyInclusiveCatalogingHistories2024}
\bibfield{editor}{\bibinfo{person}{Amber Billey}, \bibinfo{person}{Elizabeth Nelson}, {and} \bibinfo{person}{Rebecca Uhl}} (Eds.). \bibinfo{year}{2024}\natexlab{}.
\newblock \bibinfo{booktitle}{\emph{Inclusive {{Cataloging}}: {{Histories}}, {{Context}}, and {{Reparative Approaches}}}}.
\newblock \bibinfo{publisher}{ALA Editions}, \bibinfo{address}{Chicago}.
\newblock
\showISBNx{9798892555661}


\bibitem[Bishop(1990)]%
        {bishopMirrorsWindowsSlidingGlassDoros}
\bibfield{author}{\bibinfo{person}{Rudine~Sims Bishop}.} \bibinfo{year}{1990}\natexlab{}.
\newblock \showarticletitle{Mirrors, Windows, and Sliding Glass Doors}.
\newblock \bibinfo{journal}{\emph{Perspectives: Choosing and Using Books for the Classroom}} \bibinfo{volume}{6}, \bibinfo{number}{3} (\bibinfo{year}{1990}).
\newblock
\urldef\tempurl%
\url{https://scenicregional.org/wp-content/uploads/2017/08/Mirrors-Windows-and-Sliding-Glass-Doors.pdf}
\showURL{%
\tempurl}


\bibitem[Borradaile et~al\mbox{.}({[n.\,d.]})]%
        {borradaileWhoseTweetsAre2020}
\bibfield{author}{\bibinfo{person}{Glencora Borradaile}, \bibinfo{person}{Brett Burkhardt}, {and} \bibinfo{person}{Alexandria LeClerc}.} \bibinfo{year}{[n.\,d.]}\natexlab{}.
\newblock \showarticletitle{Whose Tweets Are Surveilled for the Police: An Audit of a Social-Media Monitoring Tool via Log Files}. In \bibinfo{booktitle}{\emph{Proceedings of the 2020 {{Conference}} on {{Fairness}}, {{Accountability}}, and {{Transparency}}}} (New York, NY, USA, 2020-01-27) \emph{(\bibinfo{series}{{{FAT}}* '20})}. \bibinfo{publisher}{Association for Computing Machinery}, \bibinfo{pages}{570--580}.
\newblock
\showISBNx{978-1-4503-6936-7}
\urldef\tempurl%
\url{https://doi.org/10.1145/3351095.3372841}
\showDOI{\tempurl}


\bibitem[Breeding(2019)]%
        {OverDrivesNewOwners}
\bibfield{author}{\bibinfo{person}{Marhsall Breeding}.} \bibinfo{year}{2019}\natexlab{}.
\newblock \bibinfo{booktitle}{\emph{{{OverDrive}}’s {{New Owners}}: {{What It Means}}}}.
\newblock American Libraries Magazine.
\newblock
\urldef\tempurl%
\url{https://americanlibrariesmagazine.org/blogs/the-scoop/overdrives-new-owners-what-means/}
\showURL{%
\tempurl}
\newblock
\shownote{Accessed: 2025-01-22}.


\bibitem[Buolamwini and Gebru(2018)]%
        {buolamwiniGenderShadesIntersectional2018}
\bibfield{author}{\bibinfo{person}{Joy Buolamwini} {and} \bibinfo{person}{Timnit Gebru}.} \bibinfo{year}{2018}\natexlab{}.
\newblock \showarticletitle{Gender {{Shades}}: {{Intersectional Accuracy Disparities}} in {{Commercial Gender Classification}}}. In \bibinfo{booktitle}{\emph{Proceedings of the 1st {{Conference}} on {{Fairness}}, {{Accountability}} and {{Transparency}}}}. \bibinfo{publisher}{PMLR}, \bibinfo{pages}{77--91}.
\newblock
\showISSN{2640-3498}


\bibitem[Bureau(2021)]%
        {bureau2020CensusIlluminates}
\bibfield{author}{\bibinfo{person}{US~Census Bureau}.} \bibinfo{year}{2021}\natexlab{}.
\newblock \bibinfo{title}{2020 {{Census Illuminates Racial}} and {{Ethnic Composition}} of the {{Country}}}.
\newblock \bibinfo{howpublished}{https://www.census.gov/library/stories/2021/08/improved-race-ethnicity-measures-reveal-united-states-population-much-more-multiracial.html}.
\newblock


\bibitem[Chadley(1992)]%
        {chadley1992addressing}
\bibfield{author}{\bibinfo{person}{Otis~A Chadley}.} \bibinfo{year}{1992}\natexlab{}.
\newblock \showarticletitle{Addressing cultural diversity in academic and research libraries}.
\newblock \bibinfo{journal}{\emph{College \& research libraries}} \bibinfo{volume}{53}, \bibinfo{number}{3} (\bibinfo{year}{1992}), \bibinfo{pages}{206--214}.
\newblock
\urldef\tempurl%
\url{https://doi.org/10.5860/crl_53_03_206}
\showURL{%
\tempurl}


\bibitem[Chang(2023)]%
        {changQueerGapCultural2023}
\bibfield{author}{\bibinfo{person}{Kent~K. Chang}.} \bibinfo{year}{2023}\natexlab{}.
\newblock \showarticletitle{The {{Queer Gap}} in {{Cultural Analytics}}}.
\newblock \bibinfo{journal}{\emph{Debates in the Digital Humanities 2023}} (\bibinfo{year}{2023}), \bibinfo{pages}{105--119}.
\newblock


\bibitem[Chang et~al\mbox{.}(2024)]%
        {changSubversiveCharactersStereotyping2024}
\bibfield{author}{\bibinfo{person}{Kent~K. Chang}, \bibinfo{person}{Anna Ho}, {and} \bibinfo{person}{David Bamman}.} \bibinfo{year}{2024}\natexlab{}.
\newblock \showarticletitle{Subversive {{Characters}} and {{Stereotyping Readers}}: {{Characterizing Queer Relationalities}} with {{Dialogue-Based Relation Extraction}}}. In \bibinfo{booktitle}{\emph{Computational {{Humanities Research}}}}. \bibinfo{publisher}{arXiv}.
\newblock
\urldef\tempurl%
\url{https://doi.org/10.48550/arXiv.2410.14978}
\showDOI{\tempurl}
\showeprint[arxiv]{2410.14978}~[cs]


\bibitem[Chun(2009)]%
        {chunIntroductionRaceTechnology2009}
\bibfield{author}{\bibinfo{person}{Wendy Hui~Kyong Chun}.} \bibinfo{year}{2009}\natexlab{}.
\newblock \showarticletitle{Introduction: {{Race}} and/as {{Technology}}; or, {{How}} to {{Do Things}} to {{Race}}}.
\newblock \bibinfo{journal}{\emph{Camera Obscura: Feminism, Culture, and Media Studies}} \bibinfo{volume}{24}, \bibinfo{number}{1 (70)} (\bibinfo{date}{May} \bibinfo{year}{2009}), \bibinfo{pages}{7--35}.
\newblock
\showISSN{0270-5346}
\urldef\tempurl%
\url{https://doi.org/10.1215/02705346-2008-013}
\showDOI{\tempurl}


\bibitem[Ciszek and Young(2010)]%
        {ciszek2010diversity}
\bibfield{author}{\bibinfo{person}{Matthew~P Ciszek} {and} \bibinfo{person}{Courtney~L Young}.} \bibinfo{year}{2010}\natexlab{}.
\newblock \showarticletitle{Diversity collection assessment in large academic libraries}.
\newblock \bibinfo{journal}{\emph{Collection Building}} \bibinfo{volume}{29}, \bibinfo{number}{4} (\bibinfo{year}{2010}), \bibinfo{pages}{154--161}.
\newblock
\urldef\tempurl%
\url{https://doi.org/10.1108/01604951011088899}
\showURL{%
\tempurl}


\bibitem[Clarke and Braun(2017)]%
        {clarkeThematicAnalysis2017}
\bibfield{author}{\bibinfo{person}{Victoria Clarke} {and} \bibinfo{person}{Virginia Braun}.} \bibinfo{year}{2017}\natexlab{}.
\newblock \showarticletitle{Thematic Analysis}.
\newblock \bibinfo{journal}{\emph{The Journal of Positive Psychology}} \bibinfo{volume}{12}, \bibinfo{number}{3} (\bibinfo{date}{May} \bibinfo{year}{2017}), \bibinfo{pages}{297--298}.
\newblock
\showISSN{1743-9760, 1743-9779}
\urldef\tempurl%
\url{https://doi.org/10.1080/17439760.2016.1262613}
\showDOI{\tempurl}


\bibitem[CollectionHQ(2025)]%
        {collectionhqDiversityEquityInclusion2025}
\bibfield{author}{\bibinfo{person}{CollectionHQ}.} \bibinfo{year}{2025}\natexlab{}.
\newblock \bibinfo{title}{Diversity, {{Equity}} and {{Inclusion Analysis}}}.
\newblock \bibinfo{howpublished}{https://www.collectionhq.com/diversity-analysis-2/}.
\newblock


\bibitem[Collins(2012)]%
        {collinsCurrentBudgetEnvironment2012}
\bibfield{author}{\bibinfo{person}{Tim Collins}.} \bibinfo{year}{2012}\natexlab{}.
\newblock \showarticletitle{The {{Current Budget Environment}} and {{Its Impact}} on {{Libraries}}, {{Publishers}} and {{Vendors}}}.
\newblock \bibinfo{journal}{\emph{Journal of Library Administration}} \bibinfo{volume}{52}, \bibinfo{number}{1} (\bibinfo{date}{Jan.} \bibinfo{year}{2012}), \bibinfo{pages}{18--35}.
\newblock
\showISSN{0193-0826}
\urldef\tempurl%
\url{https://doi.org/10.1080/01930826.2012.630643}
\showDOI{\tempurl}


\bibitem[Corrigan et~al\mbox{.}(2022)]%
        {corrigan2022banned}
\bibfield{author}{\bibinfo{person}{Jack Corrigan}, \bibinfo{person}{Sergio Fontanez}, {and} \bibinfo{person}{Michael Kratsios}.} \bibinfo{year}{2022}\natexlab{}.
\newblock \showarticletitle{Banned in DC: Examining Government Approaches to Foreign Technology Threats}.
\newblock  (\bibinfo{date}{Oct} \bibinfo{year}{2022}).
\newblock
\urldef\tempurl%
\url{https://doi.org/10.51593/20220007}
\showURL{%
\tempurl}


\bibitem[{Costanza-Chock} et~al\mbox{.}(2022)]%
        {costanza-chockWhoAuditsAuditors2022}
\bibfield{author}{\bibinfo{person}{Sasha {Costanza-Chock}}, \bibinfo{person}{Inioluwa~Deborah Raji}, {and} \bibinfo{person}{Joy Buolamwini}.} \bibinfo{year}{2022}\natexlab{}.
\newblock \showarticletitle{Who {{Audits}} the {{Auditors}}? {{Recommendations}} from a Field Scan of the Algorithmic Auditing Ecosystem}. In \bibinfo{booktitle}{\emph{2022 {{ACM Conference}} on {{Fairness}}, {{Accountability}}, and {{Transparency}}}}. \bibinfo{publisher}{ACM}, \bibinfo{address}{Seoul Republic of Korea}, \bibinfo{pages}{1571--1583}.
\newblock
\showISBNx{978-1-4503-9352-2}
\urldef\tempurl%
\url{https://doi.org/10.1145/3531146.3533213}
\showDOI{\tempurl}


\bibitem[Diaz(2023)]%
        {FloridaDEIban}
\bibfield{author}{\bibinfo{person}{Jaclyn Diaz}.} \bibinfo{year}{2023}\natexlab{}.
\newblock \showarticletitle{Florida Gov. Ron DeSantis signs a bill banning DEI initiatives in public colleges}.
\newblock \bibinfo{journal}{\emph{NPR}} (\bibinfo{date}{15 May} \bibinfo{year}{2023}).
\newblock


\bibitem[Drabinski(2013)]%
        {drabinskiQueeringCatalogQueer2013}
\bibfield{author}{\bibinfo{person}{Emily Drabinski}.} \bibinfo{year}{2013}\natexlab{}.
\newblock \showarticletitle{Queering the {{Catalog}}: {{Queer Theory}} and the {{Politics}} of {{Correction}}}.
\newblock \bibinfo{journal}{\emph{The Library Quarterly}} \bibinfo{volume}{83}, \bibinfo{number}{2} (\bibinfo{date}{April} \bibinfo{year}{2013}), \bibinfo{pages}{94--111}.
\newblock
\showISSN{0024-2519}
\urldef\tempurl%
\url{https://doi.org/10.1086/669547}
\showDOI{\tempurl}


\bibitem[Duarte and {Belarde-Lewis}(2015)]%
        {duarteImaginingCreatingSpaces2015}
\bibfield{author}{\bibinfo{person}{Marisa~Elena Duarte} {and} \bibinfo{person}{Miranda {Belarde-Lewis}}.} \bibinfo{year}{2015}\natexlab{}.
\newblock \showarticletitle{Imagining: {{Creating Spaces}} for {{Indigenous Ontologies}}}.
\newblock \bibinfo{journal}{\emph{Cataloging \& Classification Quarterly}} \bibinfo{volume}{53}, \bibinfo{number}{5-6} (\bibinfo{date}{July} \bibinfo{year}{2015}), \bibinfo{pages}{677--702}.
\newblock
\showISSN{0163-9374}
\urldef\tempurl%
\url{https://doi.org/10.1080/01639374.2015.1018396}
\showDOI{\tempurl}


\bibitem[Emerson and Lehman(2022)]%
        {EMERSON2022102517}
\bibfield{author}{\bibinfo{person}{María~Evelia Emerson} {and} \bibinfo{person}{Lauryn~Grace Lehman}.} \bibinfo{year}{2022}\natexlab{}.
\newblock \showarticletitle{Who Are We Missing? Conducting a Diversity Audit in a Liberal Arts College Library}.
\newblock \bibinfo{journal}{\emph{The Journal of Academic Librarianship}} \bibinfo{volume}{48}, \bibinfo{number}{3} (\bibinfo{year}{2022}), \bibinfo{pages}{102517}.
\newblock
\showISSN{0099-1333}
\urldef\tempurl%
\url{https://doi.org/10.1016/j.acalib.2022.102517}
\showDOI{\tempurl}


\bibitem[Enis(2014)]%
        {enisBakerTaylorCollectionHQ2014}
\bibfield{author}{\bibinfo{person}{Matt Enis}.} \bibinfo{year}{2014}\natexlab{}.
\newblock \bibinfo{title}{Baker \& {{Taylor}}, {{collectionHQ Launch ESP}}, a {{Predictive Collection Development Tool}}}.
\newblock \bibinfo{howpublished}{https://www.libraryjournal.com/story/baker-taylor-collectionhq-launch-esp-a-predictive-collection-development-tool}.
\newblock
\urldef\tempurl%
\url{https://www.libraryjournal.com/story/baker-taylor-collectionhq-launch-esp-a-predictive-collection-development-tool}
\showURL{%
\tempurl}


\bibitem[Enis(2024)]%
        {enisAIHorizon2024}
\bibfield{author}{\bibinfo{person}{Matt Enis}.} \bibinfo{year}{2024}\natexlab{}.
\newblock \showarticletitle{{{AI}} on the {{Horizon}}}.
\newblock \bibinfo{journal}{\emph{Library Journal}} (\bibinfo{date}{Oct} \bibinfo{year}{2024}).
\newblock
\urldef\tempurl%
\url{https://www.libraryjournal.com/story/ai-on-the-horizon}
\showURL{%
\tempurl}
\newblock
\shownote{Accessed: 2025-01-21}.


\bibitem[Escudero(2023)]%
        {escuderoHowAlbanyAdding2023}
\bibfield{author}{\bibinfo{person}{Shayla Escudero}.} \bibinfo{year}{2023}\natexlab{}.
\newblock \showarticletitle{How {{Albany}} Is Adding Diversity to the Shelves at Its Library}.
\newblock  (\bibinfo{year}{2023}).
\newblock
\urldef\tempurl%
\url{https://democratherald.com/news/local/article_d3b2735e-deed-11ed-9cd7-c3306e7d1d76.html}
\showURL{%
\tempurl}


\bibitem[Eubanks(2018)]%
        {eubanksAutomatingInequalityHow2018}
\bibfield{author}{\bibinfo{person}{Virginia Eubanks}.} \bibinfo{year}{2018}\natexlab{}.
\newblock \bibinfo{booktitle}{\emph{Automating {{Inequality}}: {{How High-Tech Tools Profile}}, {{Police}}, and {{Punish}} the {{Poor}}}}.
\newblock \bibinfo{publisher}{St. Martin's Press}, \bibinfo{address}{New York, NY}.
\newblock
\showISBNx{978-1-250-07431-7}


\bibitem[EveryLibrary(2024)]%
        {2024BillTracking}
\bibfield{author}{\bibinfo{person}{EveryLibrary}.} \bibinfo{year}{2024}\natexlab{}.
\newblock \bibinfo{title}{Legislation of Concern in 2024}.
\newblock
\newblock
\urldef\tempurl%
\url{https://www.everylibrary.org/billtracking2024}
\showURL{%
\tempurl}


\bibitem[EveryLibrary(2025)]%
        {2025BillTracking}
\bibfield{author}{\bibinfo{person}{EveryLibrary}.} \bibinfo{year}{2025}\natexlab{}.
\newblock \bibinfo{title}{Legislation of Concern}.
\newblock
\newblock
\urldef\tempurl%
\url{https://www.everylibrary.org/billtracking}
\showURL{%
\tempurl}


\bibitem[Farmanfarmaian(1997)]%
        {farmanfarmaian1997hawaii}
\bibfield{author}{\bibinfo{person}{Roxanne Farmanfarmaian}.} \bibinfo{year}{1997}\natexlab{}.
\newblock \showarticletitle{Hawaii libraries vs. Baker \& Taylor: better times ahead?}
\newblock \bibinfo{journal}{\emph{Publishers Weekly}} \bibinfo{volume}{244}, \bibinfo{number}{9} (\bibinfo{year}{1997}), \bibinfo{pages}{16--17}.
\newblock


\bibitem[Fischer(2023)]%
        {fischerAutomatingDiversityAudit2023}
\bibfield{author}{\bibinfo{person}{Rachel Fischer}.} \bibinfo{year}{2023}\natexlab{}.
\newblock \showarticletitle{Automating the {{Diversity Audit Process}}}.
\newblock \bibinfo{journal}{\emph{Information Technology and Libraries}} \bibinfo{volume}{42}, \bibinfo{number}{3} (\bibinfo{date}{Sept.} \bibinfo{year}{2023}).
\newblock
\showISSN{2163-5226}
\urldef\tempurl%
\url{https://doi.org/10.5860/ital.v42i3.16925}
\showDOI{\tempurl}


\bibitem[for Liberty(2024)]%
        {AbooutMomsforLiberty}
\bibfield{author}{\bibinfo{person}{Moms for Liberty}.} \bibinfo{year}{2024}\natexlab{}.
\newblock \bibinfo{title}{Who We Are}.
\newblock
\newblock
\urldef\tempurl%
\url{https://www.momsforliberty.org/about/}
\showURL{%
\tempurl}


\bibitem[Freund and Weng(2024)]%
        {freundDedicatedDocketUS2024}
\bibfield{author}{\bibinfo{person}{Daniel Freund} {and} \bibinfo{person}{Wentao Weng}.} \bibinfo{year}{2024}\natexlab{}.
\newblock \showarticletitle{The {{Dedicated Docket}} in {{U}}.{{S}}. {{Immigration Courts}}: {{An}} Analysis of Fairness and Efficiency Properties}. In \bibinfo{booktitle}{\emph{Proceedings of the 25th {{ACM Conference}} on {{Economics}} and {{Computation}}}} (New York, NY, USA, 2024-12-17) \emph{(\bibinfo{series}{{{EC}} '24})}. \bibinfo{publisher}{Association for Computing Machinery}, \bibinfo{pages}{346}.
\newblock
\showISBNx{9798400707049}
\urldef\tempurl%
\url{https://doi.org/10.1145/3670865.3673541}
\showDOI{\tempurl}


\bibitem[{Fuller-Gregory}(2022)]%
        {fuller-gregoryAuditsWholePicture}
\bibfield{author}{\bibinfo{person}{Christina {Fuller-Gregory}}.} \bibinfo{year}{2022}\natexlab{}.
\newblock \bibinfo{title}{{{DEI Audits}}: {{The Whole Picture}} {\textbar} {{Equity}}}.
\newblock \bibinfo{howpublished}{https://www.libraryjournal.com/story/DEI-Audits-The-Whole-Picture-Equity}.
\newblock
\newblock
\shownote{Accessed: 2025-01-14}.


\bibitem[Gallon(2016)]%
        {gallonMakingCaseBlack2016}
\bibfield{author}{\bibinfo{person}{Kim Gallon}.} \bibinfo{year}{2016}\natexlab{}.
\newblock \showarticletitle{Making a Case for the Black Digital Humanities}.
\newblock In \bibinfo{booktitle}{\emph{Debates in the {{Digital Humanities}} 2016}}, \bibfield{editor}{\bibinfo{person}{Matthew~K. Gold} {and} \bibinfo{person}{Lauren~F. Klein}} (Eds.). \bibinfo{publisher}{University of Minnesota Press}.
\newblock
\showISBNx{978-1-4529-5148-5 978-0-8166-9954-4}
\urldef\tempurl%
\url{https://doi.org/10.5749/j.ctt1cn6thb}
\showDOI{\tempurl}
\showeprint[jstor]{10.5749/j.ctt1cn6thb}


\bibitem[Gardiner(2025)]%
        {gardinerJudgeBlocksTrump2025}
\bibfield{author}{\bibinfo{person}{Dustin Gardiner}.} \bibinfo{year}{2025}\natexlab{}.
\newblock \bibinfo{title}{Judge Blocks {{Trump}} Admin from Pulling {{San Francisco}} Housing Funds over Anti-{{DEI}} Rules}.
\newblock \bibinfo{howpublished}{https://www.politico.com/news/2025/05/08/trump-san-francisco-housing-dei-00336454}.
\newblock


\bibitem[Gates et~al\mbox{.}(2022)]%
        {gatesUsingDataCollection2022}
\bibfield{author}{\bibinfo{person}{Anitra Gates}, \bibinfo{person}{Amberlee McGaughey}, \bibinfo{person}{Celia Mulder}, {and} \bibinfo{person}{Sarah Voels}.} \bibinfo{year}{2022}\natexlab{}.
\newblock \showarticletitle{Using {{Data From Collection Diversity Audits}}}.
\newblock \bibinfo{journal}{\emph{Computers in Libraries}} \bibinfo{volume}{42}, \bibinfo{number}{9} (\bibinfo{date}{Nov.} \bibinfo{year}{2022}).
\newblock


\bibitem[Groves et~al\mbox{.}(2024)]%
        {grovesAuditingWorkExploring2024}
\bibfield{author}{\bibinfo{person}{Lara Groves}, \bibinfo{person}{Jacob Metcalf}, \bibinfo{person}{Alayna Kennedy}, \bibinfo{person}{Briana Vecchione}, {and} \bibinfo{person}{Andrew Strait}.} \bibinfo{year}{2024}\natexlab{}.
\newblock \showarticletitle{Auditing {{Work}}: {{Exploring}} the {{New York City}} Algorithmic Bias Audit Regime}. In \bibinfo{booktitle}{\emph{Proceedings of the 2024 {{ACM Conference}} on {{Fairness}}, {{Accountability}}, and {{Transparency}}}} \emph{(\bibinfo{series}{{{FAccT}} '24})}. \bibinfo{publisher}{Association for Computing Machinery}, \bibinfo{address}{New York, NY, USA}, \bibinfo{pages}{1107--1120}.
\newblock
\showISBNx{9798400704505}
\urldef\tempurl%
\url{https://doi.org/10.1145/3630106.3658959}
\showDOI{\tempurl}


\bibitem[Hanna et~al\mbox{.}({[n.\,d.]})]%
        {hannaCriticalRaceMethodology2020}
\bibfield{author}{\bibinfo{person}{Alex Hanna}, \bibinfo{person}{Emily Denton}, \bibinfo{person}{Andrew Smart}, {and} \bibinfo{person}{Jamila Smith-Loud}.} \bibinfo{year}{[n.\,d.]}\natexlab{}.
\newblock \showarticletitle{Towards a Critical Race Methodology in Algorithmic Fairness}. In \bibinfo{booktitle}{\emph{Proceedings of the 2020 {{Conference}} on {{Fairness}}, {{Accountability}}, and {{Transparency}}}} (New York, NY, USA, 2020-01-27) \emph{(\bibinfo{series}{{{FAT}}* '20})}. \bibinfo{publisher}{Association for Computing Machinery}, \bibinfo{pages}{501--512}.
\newblock
\showISBNx{978-1-4503-6936-7}
\urldef\tempurl%
\url{https://doi.org/10.1145/3351095.3372826}
\showDOI{\tempurl}


\bibitem[Hoffert(2007)]%
        {hoffert2007s}
\bibfield{author}{\bibinfo{person}{Barbara Hoffert}.} \bibinfo{year}{2007}\natexlab{}.
\newblock \showarticletitle{Who's selecting now? As Phoenix Public Library boldly passes on selection responsibilities to its vendors, some libraries follow--and others dig in}.
\newblock \bibinfo{journal}{\emph{Library Journal}} \bibinfo{volume}{132}, \bibinfo{number}{14} (\bibinfo{year}{2007}), \bibinfo{pages}{40--44}.
\newblock


\bibitem[House(2025a)]%
        {EO4}
\bibfield{author}{\bibinfo{person}{The~White House}.} \bibinfo{year}{2025}\natexlab{a}.
\newblock \showarticletitle{Continuing the Reduction of the Federal Bureaucracy}.
\newblock  (\bibinfo{date}{14 March} \bibinfo{year}{2025}).
\newblock
\urldef\tempurl%
\url{https://www.whitehouse.gov/presidential-actions/2025/03/continuing-the-reduction-of-the-federal-bureaucracy/}
\showURL{%
\tempurl}


\bibitem[House(2025b)]%
        {thewhitehouseEndingRadicalWasteful2025}
\bibfield{author}{\bibinfo{person}{The~White House}.} \bibinfo{year}{2025}\natexlab{b}.
\newblock \bibinfo{title}{Ending {{Radical And Wasteful Government DEI Programs And Preferencing}}}.
\newblock \bibinfo{howpublished}{https://www.whitehouse.gov/presidential-actions/2025/01/ending-radical-and-wasteful-government-dei-programs-and-preferencing/}.
\newblock


\bibitem[Huang and Liem(2022)]%
        {huangSocialInclusionCurated2022}
\bibfield{author}{\bibinfo{person}{Han-Yin Huang} {and} \bibinfo{person}{Cynthia C.~S. Liem}.} \bibinfo{year}{2022}\natexlab{}.
\newblock \showarticletitle{Social {{Inclusion}} in {{Curated Contexts}}: {{Insights}} from {{Museum Practices}}}. In \bibinfo{booktitle}{\emph{Proceedings of the 2022 {{ACM Conference}} on {{Fairness}}, {{Accountability}}, and {{Transparency}}}} \emph{(\bibinfo{series}{{{FAccT}} '22})}. \bibinfo{publisher}{Association for Computing Machinery}, \bibinfo{address}{New York, NY, USA}, \bibinfo{pages}{300--309}.
\newblock
\showISBNx{978-1-4503-9352-2}
\urldef\tempurl%
\url{https://doi.org/10.1145/3531146.3533095}
\showDOI{\tempurl}


\bibitem[Ingram(2021)]%
        {ingramIngramAnnouncesNew2021}
\bibfield{author}{\bibinfo{person}{Ingram}.} \bibinfo{year}{2021}\natexlab{}.
\newblock \bibinfo{booktitle}{\emph{Ingram {{Announces A New Service}} to {{Help Libraries Effectively Diversify}} Their {{Print Collection}}}}.
\newblock
\urldef\tempurl%
\url{https://web.archive.org/web/20250429195151/https://www.ingramcontent.com/news/ingram-content-group-announces-a-new-service-to-help-libraries-effectively-diversify-their-print-col}
\showURL{%
\tempurl}


\bibitem[Ingram(2025a)]%
        {ingramICurate2025}
\bibfield{author}{\bibinfo{person}{Ingram}.} \bibinfo{year}{2025}\natexlab{a}.
\newblock \bibinfo{title}{{{iCurate}}}.
\newblock \bibinfo{howpublished}{https://www.ingramcontent.com/libraries/collection-development/icurate}.
\newblock


\bibitem[Ingram(2025b)]%
        {ingramICurateInClusiveSchools2025}
\bibfield{author}{\bibinfo{person}{Ingram}.} \bibinfo{year}{2025}\natexlab{b}.
\newblock \bibinfo{booktitle}{\emph{{{iCurate inClusive}} for {{Schools Frequently Asked Questions}}}}.
\newblock
\urldef\tempurl%
\url{https://web.archive.org/web/20250429171336/https://www.ingramcontent.com/libraries-page/icurate-inclusive-schools-faqs}
\showURL{%
\tempurl}


\bibitem[Jensen(2018)]%
        {jensenDiversityAuditing101}
\bibfield{author}{\bibinfo{person}{Karen Jensen}.} \bibinfo{year}{2018}\natexlab{}.
\newblock \bibinfo{booktitle}{\emph{Diversity {{Auditing}} 101: {{How}} to {{Evaluate Your Collection}}}}.
\newblock School Library Journal.
\newblock
\urldef\tempurl%
\url{https://www.slj.com/story/diversity-auditing-101-how-to-evaluate-collection}
\showURL{%
\tempurl}


\bibitem[Jensen and {comments}(2017)]%
        {jensenDoingYACollection2017}
\bibfield{author}{\bibinfo{person}{Karen Jensen} {and} \bibinfo{person}{MLS~8 {comments}}.} \bibinfo{year}{2017}\natexlab{}.
\newblock \bibinfo{booktitle}{\emph{Doing a {{YA Collection Diversity Audit}}: {{Understanding Your Local Community}} ({{Part}} 1)}}.
\newblock Teen Librarian Toolbox.
\newblock
\urldef\tempurl%
\url{https://teenlibrariantoolbox.com/2017/11/01/doing-a-diversity-audit-understanding-your-local-community/}
\showURL{%
\tempurl}


\bibitem[Jo and Gebru(2020)]%
        {joLessonsArchivesStrategies2020}
\bibfield{author}{\bibinfo{person}{Eun~Seo Jo} {and} \bibinfo{person}{Timnit Gebru}.} \bibinfo{year}{2020}\natexlab{}.
\newblock \showarticletitle{Lessons from Archives: Strategies for Collecting Sociocultural Data in Machine Learning}. In \bibinfo{booktitle}{\emph{Proceedings of the 2020 {{Conference}} on {{Fairness}}, {{Accountability}}, and {{Transparency}}}} \emph{(\bibinfo{series}{{{FAT}}* '20})}. \bibinfo{publisher}{Association for Computing Machinery}, \bibinfo{address}{New York, NY, USA}, \bibinfo{pages}{306--316}.
\newblock
\showISBNx{978-1-4503-6936-7}
\urldef\tempurl%
\url{https://doi.org/10.1145/3351095.3372829}
\showDOI{\tempurl}


\bibitem[Kemp(2008)]%
        {kempMARCRecordServices2008}
\bibfield{author}{\bibinfo{person}{Rebecca Kemp}.} \bibinfo{year}{2008}\natexlab{}.
\newblock \showarticletitle{{{MARC Record Services}}: {{A Comparative Study}} of {{Library Practices}} and {{Perceptions}}}.
\newblock \bibinfo{journal}{\emph{The Serials Librarian}} \bibinfo{volume}{55}, \bibinfo{number}{3} (\bibinfo{date}{Sept.} \bibinfo{year}{2008}), \bibinfo{pages}{379--410}.
\newblock
\showISSN{0361-526X}
\urldef\tempurl%
\url{https://doi.org/10.1080/03615260802056367}
\showDOI{\tempurl}


\bibitem[Kilgour(1970)]%
        {kilgourHistoryLibraryComputerization1970}
\bibfield{author}{\bibinfo{person}{Frederick~G. Kilgour}.} \bibinfo{year}{1970}\natexlab{}.
\newblock \showarticletitle{History of {{Library Computerization}}}.
\newblock \bibinfo{journal}{\emph{Information Technology and Libraries}} \bibinfo{volume}{3}, \bibinfo{number}{3} (\bibinfo{date}{Sept.} \bibinfo{year}{1970}), \bibinfo{pages}{218--229}.
\newblock
\showISSN{2163-5226}
\urldef\tempurl%
\url{https://doi.org/10.6017/ital.v3i3.5256}
\showDOI{\tempurl}


\bibitem[Knuth and Bair-Mundy(1998)]%
        {knuth1998revolt}
\bibfield{author}{\bibinfo{person}{Rebecca Knuth} {and} \bibinfo{person}{Donna~G Bair-Mundy}.} \bibinfo{year}{1998}\natexlab{}.
\newblock \showarticletitle{Revolt over outsourcing: Hawaii's librarians speak out about contracted selection}.
\newblock \bibinfo{journal}{\emph{Collection management}} \bibinfo{volume}{23}, \bibinfo{number}{1-2} (\bibinfo{year}{1998}), \bibinfo{pages}{81--112}.
\newblock
\urldef\tempurl%
\url{https://doi.org/10.1300/J105v23n01_04}
\showURL{%
\tempurl}


\bibitem[Lab(2025)]%
        {cataloginglabProblemLCSH2025}
\bibfield{author}{\bibinfo{person}{Cataloging Lab}.} \bibinfo{year}{2025}\natexlab{}.
\newblock \bibinfo{title}{Problem {{LCSH}}}.
\newblock
\newblock


\bibitem[Lamdan(2022)]%
        {lamdanDataCartelsCompanies2022}
\bibfield{author}{\bibinfo{person}{Sarah Lamdan}.} \bibinfo{year}{2022}\natexlab{}.
\newblock \bibinfo{booktitle}{\emph{Data {{Cartels}}: {{The Companies That Control}} and {{Monopolize Our Information}}}}.
\newblock \bibinfo{publisher}{Stanford University Press}.
\newblock
\showISBNx{978-1-5036-3371-1}


\bibitem[Larrick(1965)]%
        {larrickAllwhiteWorldChildrens1965}
\bibfield{author}{\bibinfo{person}{Nancy Larrick}.} \bibinfo{year}{1965}\natexlab{}.
\newblock \showarticletitle{The All-White World of Children’s Books}.
\newblock  \bibinfo{volume}{48}, \bibinfo{number}{11} (\bibinfo{year}{1965}), \bibinfo{pages}{63--65}.
\newblock


\bibitem[Lawrence(1997)]%
        {lawrenceEachOthersHarvest1997}
\bibfield{author}{\bibinfo{person}{Charles R.~III Lawrence}.} \bibinfo{year}{1997}\natexlab{}.
\newblock \showarticletitle{Each {{Other}}'s {{Harvest}}: {{Diversity}}'s {{Deeper Meaning}}}.
\newblock \bibinfo{journal}{\emph{University of San Francisco Law Review}} (\bibinfo{year}{1997}).
\newblock
\urldef\tempurl%
\url{http://hdl.handle.net/10125/65961}
\showURL{%
\tempurl}


\bibitem[Leong(2025)]%
        {leongDiversityMessagingAffirmative2025a}
\bibfield{author}{\bibinfo{person}{Nancy Leong}.} \bibinfo{year}{2025}\natexlab{}.
\newblock \showarticletitle{Diversity {{Messaging After Affirmative Action}}}.
\newblock \bibinfo{journal}{\emph{Minnesota Law Review}} (\bibinfo{date}{Feb.} \bibinfo{year}{2025}).
\newblock


\bibitem[Libraries(2025)]%
        {canadianschoollibrariesDiversityAudits2025}
\bibfield{author}{\bibinfo{person}{Canadian~School Libraries}.} \bibinfo{year}{2025}\natexlab{}.
\newblock \bibinfo{booktitle}{\emph{Diversity {{Audits}}}}.
\newblock
\urldef\tempurl%
\url{https://www.canadianschoollibraries.ca/collection-diversity-toolkit/diversity-audits/}
\showURL{%
\tempurl}


\bibitem[Library(2023)]%
        {theseattlepubliclibraryBoardTrusteesMeeting2023}
\bibfield{author}{\bibinfo{person}{The Seattle~Public Library}.} \bibinfo{year}{2023}\natexlab{}.
\newblock \bibinfo{title}{Board of {{Trustees Meeting}}}.
\newblock
\newblock
\urldef\tempurl%
\url{https://www.spl.org/Seattle-Public-Library/documents/about-us/board-meetings/2023%20library%20board%20meeting%20materials/2023%20July%2027%20Library%20Board%20Meeting%20Packet.pdf}
\showURL{%
\tempurl}


\bibitem[{libraryIQ}(2025)]%
        {libraryiqDiversity2025}
\bibfield{author}{\bibinfo{person}{{libraryIQ}}.} \bibinfo{year}{2025}\natexlab{}.
\newblock \bibinfo{title}{Diversity}.
\newblock \bibinfo{howpublished}{https://www.libraryiq.com/diversity}.
\newblock


\bibitem[Limbong(2025a)]%
        {IMLS_cuts}
\bibfield{author}{\bibinfo{person}{Andrew Limbong}.} \bibinfo{year}{2025}\natexlab{a}.
\newblock \showarticletitle{Entire staff at federal agency that funds libraries and museums put on leave}.
\newblock \bibinfo{journal}{\emph{National Public Radio}} (\bibinfo{date}{31 March} \bibinfo{year}{2025}).
\newblock
\urldef\tempurl%
\url{https://www.npr.org/2025/03/31/nx-s1-5334415/doge-institute-of-museum-and-library-services}
\showURL{%
\tempurl}


\bibitem[Limbong(2025b)]%
        {limbongLibrarianCongressFiring2025}
\bibfield{author}{\bibinfo{person}{Andrew Limbong}.} \bibinfo{year}{2025}\natexlab{b}.
\newblock \showarticletitle{Librarian of {{Congress}} Firing Is Latest Move in Upheaval of {{U}}.{{S}}. Cultural Institutions}.
\newblock \bibinfo{journal}{\emph{NPR}} (\bibinfo{date}{May} \bibinfo{year}{2025}).
\newblock
\urldef\tempurl%
\url{https://www.npr.org/2025/05/09/nx-s1-5393737/carla-hayden-fired-library-of-congress-trump}
\showURL{%
\tempurl}


\bibitem[Littletree and {and Metoyer}(2015)]%
        {littletreeKnowledgeOrganizationIndigenous2015}
\bibfield{author}{\bibinfo{person}{Sandra Littletree} {and} \bibinfo{person}{Cheryl~A. {and Metoyer}}.} \bibinfo{year}{2015}\natexlab{}.
\newblock \showarticletitle{Knowledge {{Organization}} from an {{Indigenous Perspective}}: {{The Mashantucket Pequot Thesaurus}} of {{American Indian Terminology Project}}}.
\newblock \bibinfo{journal}{\emph{Cataloging \& Classification Quarterly}} \bibinfo{volume}{53}, \bibinfo{number}{5-6} (\bibinfo{date}{July} \bibinfo{year}{2015}), \bibinfo{pages}{640--657}.
\newblock
\showISSN{0163-9374}
\urldef\tempurl%
\url{https://doi.org/10.1080/01639374.2015.1010113}
\showDOI{\tempurl}


\bibitem[Low(2023)]%
        {lee&lowDiversityBaselineSurvey}
\bibfield{author}{\bibinfo{person}{Lee~\& Low}.} \bibinfo{year}{2023}\natexlab{}.
\newblock \bibinfo{title}{Diversity {{Baseline Survey}} 3.0}.
\newblock
\newblock
\urldef\tempurl%
\url{https://www.leeandlow.com/about/diversity-baseline-survey/dbs3/}
\showURL{%
\tempurl}


\bibitem[Mackin(2025)]%
        {mackinTagReportAnalyze2025}
\bibfield{author}{\bibinfo{person}{Mackin}.} \bibinfo{year}{2025}\natexlab{}.
\newblock \bibinfo{title}{Tag {{Report}}: {{Analyze}} \& {{Diversify Your Library Collection}}}.
\newblock
\newblock
\urldef\tempurl%
\url{https://web.archive.org/web/20250429203910/https://home.mackin.com/library/collection-management/tag-report/}
\showURL{%
\tempurl}


\bibitem[Mahomed et~al\mbox{.}(2024)]%
        {mahomedAuditingGPTsContent2024}
\bibfield{author}{\bibinfo{person}{Yaaseen Mahomed}, \bibinfo{person}{Charlie~M. Crawford}, \bibinfo{person}{Sanjana Gautam}, \bibinfo{person}{Sorelle~A. Friedler}, {and} \bibinfo{person}{Danaë Metaxa}.} \bibinfo{year}{2024}\natexlab{}.
\newblock \showarticletitle{Auditing {{GPT}}'s {{Content Moderation Guardrails}}: {{Can ChatGPT Write Your Favorite TV Show}}?}. In \bibinfo{booktitle}{\emph{Proceedings of the 2024 {{ACM Conference}} on {{Fairness}}, {{Accountability}}, and {{Transparency}}}} (New York, NY, USA, 2024-06-05) \emph{(\bibinfo{series}{{{FAccT}} '24})}. \bibinfo{publisher}{Association for Computing Machinery}, \bibinfo{pages}{660--686}.
\newblock
\showISBNx{9798400704505}
\urldef\tempurl%
\url{https://doi.org/10.1145/3630106.3658932}
\showDOI{\tempurl}


\bibitem[Malespina(2024)]%
        {malespinaAILibraryReaders2024}
\bibfield{author}{\bibinfo{person}{Elissa Malespina}.} \bibinfo{year}{2024}\natexlab{}.
\newblock \showarticletitle{{{AI}} for the {{Library}}: {{Readers}}}.
\newblock \bibinfo{journal}{\emph{School Library Journal}} (\bibinfo{date}{July} \bibinfo{year}{2024}).
\newblock
\urldef\tempurl%
\url{https://www.slj.com/story/ai-for-the-library-readers-advisory-tasks}
\showURL{%
\tempurl}


\bibitem[Mandell(2019)]%
        {mandellGenderCulturalAnalytics2019}
\bibfield{author}{\bibinfo{person}{Laura Mandell}.} \bibinfo{year}{2019}\natexlab{}.
\newblock \showarticletitle{Gender and Cultural Analytics: Finding or Making Stereotypes?}
\newblock \bibinfo{journal}{\emph{Debates in the Digital Humanities 2019}} (\bibinfo{year}{2019}), \bibinfo{pages}{3--26}.
\newblock


\bibitem[Martin and Mundle(2010)]%
        {martinCatalogingEBooksVendor2010}
\bibfield{author}{\bibinfo{person}{Kristin~E. Martin} {and} \bibinfo{person}{Kavita Mundle}.} \bibinfo{year}{2010}\natexlab{}.
\newblock \showarticletitle{Cataloging {{E-Books}} and {{Vendor Records}}}.
\newblock \bibinfo{journal}{\emph{Library Resources \& Technical Services}} \bibinfo{volume}{54}, \bibinfo{number}{4} (\bibinfo{year}{2010}), \bibinfo{pages}{227--237}.
\newblock
\showISSN{2159-9610}
\urldef\tempurl%
\url{https://doi.org/10.5860/lrts.54n4.227}
\showDOI{\tempurl}


\bibitem[Matthews(1982)]%
        {matthewsAutomatedCirculationMarketplace1982}
\bibfield{author}{\bibinfo{person}{Joseph~R. Matthews}.} \bibinfo{year}{1982}\natexlab{}.
\newblock \showarticletitle{The {{Automated Circulation Marketplace}}: {{Active}} and {{Heating Up}}}.
\newblock \bibinfo{journal}{\emph{Library Journal}} \bibinfo{volume}{107}, \bibinfo{number}{3} (\bibinfo{date}{Feb.} \bibinfo{year}{1982}), \bibinfo{pages}{233--35}.
\newblock


\bibitem[McPherson(2012)]%
        {mcphersonWhyAreDigital2012}
\bibfield{author}{\bibinfo{person}{{\relax Tara} McPherson}.} \bibinfo{year}{2012}\natexlab{}.
\newblock \showarticletitle{Why {{Are}} the {{Digital Humanities So White}}? Or {{Thinking}} the {{Histories}} of {{Race}} and {{Computation}}}.
\newblock In \bibinfo{booktitle}{\emph{Debates in the {{Digital Humanities}}}}, \bibfield{editor}{\bibinfo{person}{Matthew~K. Gold} {and} \bibinfo{person}{Lauren~F. Klein}} (Eds.). \bibinfo{publisher}{University of Minnesota Press}, \bibinfo{pages}{139--160}.
\newblock
\showISBNx{978-0-8166-7794-8}
\showeprint[jstor]{10.5749/j.ctttv8hq.12}


\bibitem[Melamed(2011)]%
        {melamedRepresentDestroyRationalizing2011}
\bibfield{author}{\bibinfo{person}{Jodi Melamed}.} \bibinfo{year}{2011}\natexlab{}.
\newblock \bibinfo{booktitle}{\emph{Represent and {{Destroy}}: {{Rationalizing Violence}} in the {{New Racial Capitalism}}}}.
\newblock \bibinfo{publisher}{Univ Of Minnesota Press}.
\newblock


\bibitem[Melamed(2020)]%
        {melamedDiversity2020}
\bibfield{author}{\bibinfo{person}{Jodi Melamed}.} \bibinfo{year}{2020}\natexlab{}.
\newblock \showarticletitle{Diversity}.
\newblock In \bibinfo{booktitle}{\emph{Keywords for {{American Cultural Studies}}}}. \bibinfo{publisher}{New York University Press}, \bibinfo{pages}{93--97}.
\newblock
\showISBNx{978-1-4798-6745-5}
\urldef\tempurl%
\url{https://doi.org/10.18574/nyu/9781479867455.003.0027}
\showURL{%
\tempurl}


\bibitem[Mortensen(2019)]%
        {mortensen2019measuring}
\bibfield{author}{\bibinfo{person}{Annabelle Mortensen}.} \bibinfo{year}{2019}\natexlab{}.
\newblock \showarticletitle{Measuring diversity in the collection}.
\newblock \bibinfo{journal}{\emph{Library Journal}} \bibinfo{volume}{144}, \bibinfo{number}{4} (\bibinfo{year}{2019}), \bibinfo{pages}{28--30}.
\newblock
\urldef\tempurl%
\url{https://www.libraryjournal.com/story/Measuring-Diversity-in-the-Collection}
\showURL{%
\tempurl}


\bibitem[Nguyen and Scribner(2025)]%
        {nguyenTrumpFiresLibrary2025}
\bibfield{author}{\bibinfo{person}{Sophia Nguyen} {and} \bibinfo{person}{Herb Scribner}.} \bibinfo{year}{2025}\natexlab{}.
\newblock \showarticletitle{Trump Fires {{Library}} of {{Congress}} Chief {{Carla Hayden}}, Citing {{DEI}}}.
\newblock \bibinfo{journal}{\emph{The Washington Post}} (\bibinfo{date}{May} \bibinfo{year}{2025}).
\newblock
\showISSN{0190-8286}


\bibitem[Noble(2018)]%
        {nobleAlgorithmsOppressionHow2018}
\bibfield{author}{\bibinfo{person}{Safiya~Umoja Noble}.} \bibinfo{year}{2018}\natexlab{}.
\newblock \bibinfo{booktitle}{\emph{Algorithms of {{Oppression}}: {{How Search Engines Reinforce Racism}}}}.
\newblock \bibinfo{publisher}{NYU Press}, \bibinfo{address}{New York}.
\newblock
\showISBNx{978-1-4798-3724-3}


\bibitem[Obermeyer and Mullainathan({[n.\,d.]})]%
        {obermeyerDissectingRacialBias2019}
\bibfield{author}{\bibinfo{person}{Ziad Obermeyer} {and} \bibinfo{person}{Sendhil Mullainathan}.} \bibinfo{year}{[n.\,d.]}\natexlab{}.
\newblock \showarticletitle{Dissecting {{Racial Bias}} in an {{Algorithm}} That {{Guides Health Decisions}} for 70 {{Million People}}}. In \bibinfo{booktitle}{\emph{Proceedings of the {{Conference}} on {{Fairness}}, {{Accountability}}, and {{Transparency}}}} (New York, NY, USA, 2019-01-29) \emph{(\bibinfo{series}{{{FAT}}* '19})}. \bibinfo{publisher}{Association for Computing Machinery}, \bibinfo{pages}{89}.
\newblock
\showISBNx{978-1-4503-6125-5}
\urldef\tempurl%
\url{https://doi.org/10.1145/3287560.3287593}
\showDOI{\tempurl}


\bibitem[OverDrive(2025)]%
        {overdriveDiversityAuditsOverDrive2025}
\bibfield{author}{\bibinfo{person}{OverDrive}.} \bibinfo{year}{2025}\natexlab{}.
\newblock \bibinfo{title}{Diversity {{Audits}} - {{OverDrive Resource Center}}}.
\newblock \bibinfo{howpublished}{https://resources.overdrive.com/library/libby-features/diversity-audit/}.
\newblock


\bibitem[Pedersen(2022)]%
        {pedersenMeasuringCollectionDiversity2022}
\bibfield{author}{\bibinfo{person}{Jordan Pedersen}.} \bibinfo{year}{2022}\natexlab{}.
\newblock \showarticletitle{Measuring {{Collection Diversity}} via {{Exploratory Analysis}} of {{Collection Metadata}}}.
\newblock \bibinfo{journal}{\emph{The Serials Librarian}} \bibinfo{volume}{82}, \bibinfo{number}{1-4} (\bibinfo{date}{May} \bibinfo{year}{2022}), \bibinfo{pages}{186--193}.
\newblock
\showISSN{0361-526X}
\urldef\tempurl%
\url{https://doi.org/10.1080/0361526X.2022.2028499}
\showDOI{\tempurl}


\bibitem[Perreault et~al\mbox{.}(2024)]%
        {perreaultAlgorithmicMisjudgementGoogle2024}
\bibfield{author}{\bibinfo{person}{Brooke Perreault}, \bibinfo{person}{Johanna~Hoonsun Lee}, \bibinfo{person}{Ropafadzo Shava}, {and} \bibinfo{person}{Eni Mustafaraj}.} \bibinfo{year}{2024}\natexlab{}.
\newblock \showarticletitle{Algorithmic {{Misjudgement}} in {{Google Search Results}}: {{Evidence}} from {{Auditing}} the {{US Online Electoral Information Environment}}}. In \bibinfo{booktitle}{\emph{Proceedings of the 2024 {{ACM Conference}} on {{Fairness}}, {{Accountability}}, and {{Transparency}}}} (New York, NY, USA, 2024-06-05) \emph{(\bibinfo{series}{{{FAccT}} '24})}. \bibinfo{publisher}{Association for Computing Machinery}, \bibinfo{pages}{433--443}.
\newblock
\showISBNx{9798400704505}
\urldef\tempurl%
\url{https://doi.org/10.1145/3630106.3658916}
\showDOI{\tempurl}


\bibitem[Price(2023)]%
        {priceNewBetaTest2023}
\bibfield{author}{\bibinfo{person}{Gary Price}.} \bibinfo{year}{2023}\natexlab{}.
\newblock \bibinfo{booktitle}{\emph{New {{Beta Test From OCLC}}: {{AI-Generated Book Recommendations}} in {{WorldCat}}.{{Org}} and {{WorldCat Find}}}}.
\newblock Library Journal infoDOCKET.
\newblock
\urldef\tempurl%
\url{https://www.infodocket.com/2023/06/21/new-beta-test-from-oclc-ai-generated-book-recommendations-in-worldcat-org-and-worldcat-find/}
\showURL{%
\tempurl}


\bibitem[Radiya-Dixit and Neff(2023)]%
        {radiya-dixitSociotechnicalAuditAssessing2023}
\bibfield{author}{\bibinfo{person}{Evani Radiya-Dixit} {and} \bibinfo{person}{Gina Neff}.} \bibinfo{year}{2023}\natexlab{}.
\newblock \showarticletitle{A {{Sociotechnical Audit}}: {{Assessing Police Use}} of {{Facial Recognition}}}. In \bibinfo{booktitle}{\emph{Proceedings of the 2023 {{ACM Conference}} on {{Fairness}}, {{Accountability}}, and {{Transparency}}}} (New York, NY, USA, 2023-06-12) \emph{(\bibinfo{series}{{{FAccT}} '23})}. \bibinfo{publisher}{Association for Computing Machinery}, \bibinfo{pages}{1334--1346}.
\newblock
\showISBNx{9798400701924}
\urldef\tempurl%
\url{https://doi.org/10.1145/3593013.3594084}
\showDOI{\tempurl}


\bibitem[Raghavan et~al\mbox{.}(2020)]%
        {ab3}
\bibfield{author}{\bibinfo{person}{Manish Raghavan}, \bibinfo{person}{Solon Barocas}, \bibinfo{person}{Jon Kleinberg}, {and} \bibinfo{person}{Karen Levy}.} \bibinfo{year}{2020}\natexlab{}.
\newblock \showarticletitle{Mitigating bias in algorithmic hiring: evaluating claims and practices}. In \bibinfo{booktitle}{\emph{Proceedings of the 2020 Conference on Fairness, Accountability, and Transparency}} (Barcelona, Spain) \emph{(\bibinfo{series}{FAT* '20})}. \bibinfo{publisher}{Association for Computing Machinery}, \bibinfo{address}{New York, NY, USA}, \bibinfo{pages}{469–481}.
\newblock
\showISBNx{9781450369367}
\urldef\tempurl%
\url{https://doi.org/10.1145/3351095.3372828}
\showDOI{\tempurl}


\bibitem[Rodriguez and Mune(2022)]%
        {rodriguezUncodingLibraryChatbots2022}
\bibfield{author}{\bibinfo{person}{Sharesly Rodriguez} {and} \bibinfo{person}{Christina Mune}.} \bibinfo{year}{2022}\natexlab{}.
\newblock \showarticletitle{Uncoding Library Chatbots: Deploying a New Virtual Reference Tool at the {{San Jose State University}} Library}.
\newblock \bibinfo{journal}{\emph{Reference Services Review}} \bibinfo{volume}{50}, \bibinfo{number}{3-4} (\bibinfo{date}{Nov.} \bibinfo{year}{2022}), \bibinfo{pages}{392--405}.
\newblock
\urldef\tempurl%
\url{https://doi.org/10.1108/RSR-05-2022-0020}
\showDOI{\tempurl}


\bibitem[Rodriguez and Mune(2025)]%
        {rodriguezLIBRARYCHATBOTSEasier2025}
\bibfield{author}{\bibinfo{person}{Sharesly Rodriguez} {and} \bibinfo{person}{Christina Mune}.} \bibinfo{year}{2025}\natexlab{}.
\newblock \showarticletitle{{{LIBRARY CHATBOTS}}: {{Easier Than You Think}}}.
\newblock \bibinfo{journal}{\emph{Computers in Libraries}} \bibinfo{volume}{41}, \bibinfo{number}{8} (\bibinfo{date}{May} \bibinfo{year}{2025}).
\newblock
\urldef\tempurl%
\url{https://www.infotoday.com/cilmag/oct21/Rodriguez-Mune--Library-Chatbots-Easier-Than-You-Think.shtml}
\showURL{%
\tempurl}


\bibitem[Rush(1988)]%
        {rushLibraryAutomationMarket1988}
\bibfield{author}{\bibinfo{person}{James Rush}.} \bibinfo{year}{1988}\natexlab{}.
\newblock \showarticletitle{The {{Library Automation Market}}: {{Why Do Vendors Fail}}? {{A History}} of {{Vendors}} and {{Their Characteristics}}}.
\newblock \bibinfo{journal}{\emph{Library Hi Tech}} \bibinfo{volume}{6}, \bibinfo{number}{3} (\bibinfo{date}{March} \bibinfo{year}{1988}), \bibinfo{pages}{7--33}.
\newblock


\bibitem[Rush(1982)]%
        {rushLibraryAutomationSystems1982}
\bibfield{author}{\bibinfo{person}{James~E. Rush}.} \bibinfo{year}{1982}\natexlab{}.
\newblock \showarticletitle{Library {{Automation Systems}} and {{Networks}}}.
\newblock In \bibinfo{booktitle}{\emph{Advances in {{Computers}}}}, \bibfield{editor}{\bibinfo{person}{Marshall~C. Yovits}} (Ed.). Vol.~\bibinfo{volume}{21}. \bibinfo{publisher}{Elsevier}, \bibinfo{pages}{333--422}.
\newblock
\urldef\tempurl%
\url{https://doi.org/10.1016/S0065-2458(08)60572-0}
\showDOI{\tempurl}


\bibitem[Schermele(2022)]%
        {schermeleCulturalPowerStruggle2022}
\bibfield{author}{\bibinfo{person}{Zachary Schermele}.} \bibinfo{year}{2022}\natexlab{}.
\newblock \showarticletitle{A Cultural Power Struggle at an {{Iowa}} Library Casts a 'dark Cloud' over a Small Town}.
\newblock \bibinfo{journal}{\emph{NBC News}} (\bibinfo{date}{Aug.} \bibinfo{year}{2022}).
\newblock
\urldef\tempurl%
\url{https://www.nbcnews.com/nbc-out/out-news/small-town-library-shut-say-culture-wars-closed-rcna39816}
\showURL{%
\tempurl}


\bibitem[Senate(2024)]%
        {GeorgiaLaw}
\bibfield{author}{\bibinfo{person}{The Georgia~State Senate}.} \bibinfo{year}{2024}\natexlab{}.
\newblock \bibinfo{title}{Senate Bill 154}.
\newblock
\newblock
\urldef\tempurl%
\url{https://www.legis.ga.gov/api/legislation/document/20232024/214457}
\showURL{%
\tempurl}


\bibitem[Sendak et~al\mbox{.}(2020)]%
        {sendakHumanBodyBlack2020}
\bibfield{author}{\bibinfo{person}{Mark Sendak}, \bibinfo{person}{Madeleine~Clare Elish}, \bibinfo{person}{Michael Gao}, \bibinfo{person}{Joseph Futoma}, \bibinfo{person}{William Ratliff}, \bibinfo{person}{Marshall Nichols}, \bibinfo{person}{Armando Bedoya}, \bibinfo{person}{Suresh Balu}, {and} \bibinfo{person}{Cara O'Brien}.} \bibinfo{year}{2020}\natexlab{}.
\newblock \showarticletitle{"{{The}} Human Body Is a Black Box": Supporting Clinical Decision-Making with Deep Learning}. In \bibinfo{booktitle}{\emph{Proceedings of the 2020 {{Conference}} on {{Fairness}}, {{Accountability}}, and {{Transparency}}}} (New York, NY, USA, 2020-01-27) \emph{(\bibinfo{series}{{{FAT}}* '20})}. \bibinfo{publisher}{Association for Computing Machinery}, \bibinfo{pages}{99--109}.
\newblock
\showISBNx{978-1-4503-6936-7}
\urldef\tempurl%
\url{https://doi.org/10.1145/3351095.3372827}
\showDOI{\tempurl}


\bibitem[Sevits and McAllister(2024)]%
        {sevits2024developing}
\bibfield{author}{\bibinfo{person}{Christi Sevits} {and} \bibinfo{person}{Alex~D McAllister}.} \bibinfo{year}{2024}\natexlab{}.
\newblock \showarticletitle{Developing an Introductory Diversity Audit at Liberty Public Library: An Action Research Approach}.
\newblock \bibinfo{journal}{\emph{Public Library Quarterly}} \bibinfo{volume}{43}, \bibinfo{number}{3} (\bibinfo{year}{2024}), \bibinfo{pages}{286--306}.
\newblock


\bibitem[Sinykin and So(2024)]%
        {Sinykin_So_2024}
\bibfield{author}{\bibinfo{person}{Dan Sinykin} {and} \bibinfo{person}{Richard~Jean So}.} \bibinfo{year}{2024}\natexlab{}.
\newblock \bibinfo{title}{Has the DEI Backlash Come for Publishing?}
\newblock
\newblock
\urldef\tempurl%
\url{https://www.theatlantic.com/books/archive/2024/06/diversity-publishing-backlash-study/678734/}
\showURL{%
\tempurl}


\bibitem[Smith(2023)]%
        {smithBuildingDiverseCollections2023a}
\bibfield{author}{\bibinfo{person}{Carrie Smith}.} \bibinfo{year}{2023}\natexlab{}.
\newblock \showarticletitle{Building {{Diverse Collections}}}.
\newblock \bibinfo{howpublished}{https://americanlibrariesmagazine.org/2023/06/01/building-diverse-collections/}.
\newblock \bibinfo{journal}{\emph{American Libraries Magazine}} (\bibinfo{date}{June} \bibinfo{year}{2023}).
\newblock


\bibitem[So and Roland(2020)]%
        {soRaceDistantReading2020}
\bibfield{author}{\bibinfo{person}{Richard~Jean So} {and} \bibinfo{person}{Edwin Roland}.} \bibinfo{year}{2020}\natexlab{}.
\newblock \showarticletitle{Race and {{Distant Reading}}}.
\newblock \bibinfo{journal}{\emph{PMLA}} \bibinfo{volume}{135}, \bibinfo{number}{1} (\bibinfo{date}{Jan.} \bibinfo{year}{2020}), \bibinfo{pages}{59--73}.
\newblock
\showISSN{0030-8129, 1938-1530}
\urldef\tempurl%
\url{https://doi.org/10.1632/pmla.2020.135.1.59}
\showDOI{\tempurl}


\bibitem[So and Wezerek(2020)]%
        {So_Wezerek_2020}
\bibfield{author}{\bibinfo{person}{Richard~Jean So} {and} \bibinfo{person}{Gus Wezerek}.} \bibinfo{year}{2020}\natexlab{}.
\newblock \showarticletitle{Just How White Is the Book Industry?}
\newblock \bibinfo{journal}{\emph{The New York Times}} (\bibinfo{date}{Dec.} \bibinfo{year}{2020}).
\newblock
\showISSN{0362-4331}
\urldef\tempurl%
\url{https://www.nytimes.com/interactive/2020/12/11/opinion/culture/diversity-publishing-industry.html}
\showURL{%
\tempurl}


\bibitem[Staff(2022a)]%
        {staffGainGreaterInsight2022}
\bibfield{author}{\bibinfo{person}{OverDrive Staff}.} \bibinfo{year}{2022}\natexlab{a}.
\newblock \bibinfo{title}{Gain Greater Insight into Your Collection with an {{OverDrive Diversity Audit}}}.
\newblock
\newblock
\urldef\tempurl%
\url{https://web.archive.org/web/20250429201829/https://company.overdrive.com/2022/06/14/gain-greater-insight-into-your-collection-with-an-overdrive-diversity-audit/}
\showURL{%
\tempurl}


\bibitem[Staff(2022b)]%
        {staffUsingFollettTitlewave2022}
\bibfield{author}{\bibinfo{person}{{\relax RILINK} Staff}.} \bibinfo{year}{2022}\natexlab{b}.
\newblock \bibinfo{title}{Using the {{Follett Titlewave Diversity Report}}: {{Session Overview}}}.
\newblock \bibinfo{howpublished}{https://guides.rilink.org/diversityreport/overview}.
\newblock


\bibitem[Swaine and Merrill(2025)]%
        {swaineAntiDEIPushNational2025}
\bibfield{author}{\bibinfo{person}{Jon Swaine} {and} \bibinfo{person}{Jeremy~B. Merrill}.} \bibinfo{year}{2025}\natexlab{}.
\newblock \showarticletitle{Amid Anti-{{DEI}} Push, {{National Park Service}} Rewrites History of {{Underground Railroad}}}.
\newblock \bibinfo{journal}{\emph{The Washington Post}} (\bibinfo{date}{April} \bibinfo{year}{2025}).
\newblock
\showISSN{0190-8286}
\urldef\tempurl%
\url{https://www.washingtonpost.com/investigations/2025/04/06/national-park-service-underground-railroad-history-slavery/}
\showURL{%
\tempurl}


\bibitem[Syme(2021)]%
        {symeAvailableNowAnalysis2021}
\bibfield{author}{\bibinfo{person}{Fiona Syme}.} \bibinfo{year}{2021}\natexlab{}.
\newblock \bibinfo{booktitle}{\emph{Available {{Now}}: {{DEI Analysis}}}}.
\newblock collectionHQ.
\newblock
\urldef\tempurl%
\url{https://web.archive.org/web/20241203103844/https://www.collectionhq.com/available-now-dei-analysis/}
\showURL{%
\tempurl}


\bibitem[Szydlowski(lish)]%
        {szydlowskiDrMartinLuther}
\bibfield{author}{\bibinfo{person}{Nick Szydlowski}.} \bibinfo{year}{2025), note = {Accessed: 2025-01-21}, abstract = {Dr. Martin Luther King, Jr. Library: Kingbot: Home}, langid = {english}}\natexlab{}.
\newblock \bibinfo{booktitle}{\emph{Dr. {{Martin Luther King}}, {{Jr}}. {{Library}}: {{Kingbot}}: {{Home}}}}.
\newblock
\urldef\tempurl%
\url{https://library.sjsu.edu/c.php?g=1425539&p=10573855}
\showURL{%
\tempurl}


\bibitem[Tape(2022)]%
        {MidwestTapeBlack2022}
\bibfield{author}{\bibinfo{person}{Midwest Tape}.} \bibinfo{year}{2022}\natexlab{}.
\newblock \bibinfo{title}{Midwest {{Tape Black Stories November}} 2022}.
\newblock
\newblock
\urldef\tempurl%
\url{https://issuu.com/midwesttape/docs/mwt\_nov2022\_blackstories\_digital/8}
\showURL{%
\tempurl}


\bibitem[Taylor(2025)]%
        {baker&taylorCollectivePurchasingAgreements2025}
\bibfield{author}{\bibinfo{person}{Baker~\& Taylor}.} \bibinfo{year}{2025}\natexlab{}.
\newblock \bibinfo{title}{Collective {{Purchasing Agreements}} - {{Detailed Information}}}.
\newblock \bibinfo{howpublished}{https://www.olservice.ca/index.php/collective-purchasing-agreements-detailed-information?product=110}.
\newblock


\bibitem[Terzis et~al\mbox{.}(2024)]%
        {terzisLawEmergingPolitical2024}
\bibfield{author}{\bibinfo{person}{Petros Terzis}, \bibinfo{person}{Michael Veale}, {and} \bibinfo{person}{Noëlle Gaumann}.} \bibinfo{year}{2024}\natexlab{}.
\newblock \showarticletitle{Law and the {{Emerging Political Economy}} of {{Algorithmic Audits}}}. In \bibinfo{booktitle}{\emph{Proceedings of the 2024 {{ACM Conference}} on {{Fairness}}, {{Accountability}}, and {{Transparency}}}} (New York, NY, USA, 2024-06-05) \emph{(\bibinfo{series}{{{FAccT}} '24})}. \bibinfo{publisher}{Association for Computing Machinery}, \bibinfo{pages}{1255--1267}.
\newblock
\showISBNx{9798400704505}
\urldef\tempurl%
\url{https://doi.org/10.1145/3630106.3658970}
\showDOI{\tempurl}


\bibitem[Vercelletto(2019)]%
        {vercellettoHowDiverseAre}
\bibfield{author}{\bibinfo{person}{Christina Vercelletto}.} \bibinfo{year}{2019}\natexlab{}.
\newblock \bibinfo{booktitle}{\emph{How {{Diverse Are Our Books}}?}}
\newblock Library Journal.
\newblock
\urldef\tempurl%
\url{https://www.libraryjournal.com/story/How-Diverse-Are-Our-Books}
\showURL{%
\tempurl}


\bibitem[Voels(2022)]%
        {voelsAuditingDiversityLibrary2022}
\bibfield{author}{\bibinfo{person}{Sarah Voels}.} \bibinfo{year}{2022}\natexlab{}.
\newblock \bibinfo{booktitle}{\emph{Auditing Diversity in Library Collections}}.
\newblock \bibinfo{publisher}{Libraries Unlimited, an imprint of ABC-CLIO, LLC}, \bibinfo{address}{Santa Barbara, California}.
\newblock
\showISBNx{978-1-4408-7875-6}


\bibitem[Vrijenhoek et~al\mbox{.}(2024)]%
        {vrijenhoekDiversityWhatDifferent2024}
\bibfield{author}{\bibinfo{person}{Sanne Vrijenhoek}, \bibinfo{person}{Savvina Daniil}, \bibinfo{person}{Jorden Sandel}, {and} \bibinfo{person}{Laura Hollink}.} \bibinfo{year}{2024}\natexlab{}.
\newblock \showarticletitle{Diversity of {{What}}? {{On}} the {{Different Conceptualizations}} of {{Diversity}} in {{Recommender Systems}}}. In \bibinfo{booktitle}{\emph{Proceedings of the 2024 {{ACM Conference}} on {{Fairness}}, {{Accountability}}, and {{Transparency}}}} \emph{(\bibinfo{series}{{{FAccT}} '24})}. \bibinfo{publisher}{Association for Computing Machinery}, \bibinfo{address}{New York, NY, USA}, \bibinfo{pages}{573--584}.
\newblock
\showISBNx{9798400704505}
\urldef\tempurl%
\url{https://doi.org/10.1145/3630106.3658926}
\showDOI{\tempurl}


\bibitem[Wallace(1997)]%
        {wallace1997outsourcing}
\bibfield{author}{\bibinfo{person}{Patricia~D Wallace}.} \bibinfo{year}{1997}\natexlab{}.
\newblock \showarticletitle{Outsourcing book selection in public and school libraries}.
\newblock \bibinfo{journal}{\emph{Collection building}} \bibinfo{volume}{16}, \bibinfo{number}{4} (\bibinfo{year}{1997}), \bibinfo{pages}{160--166}.
\newblock
\urldef\tempurl%
\url{https://doi.org/10.1108/01604959710187679}
\showDOI{\tempurl}


\bibitem[Walters(2023)]%
        {waltersAssessingDiversityAcademic2023}
\bibfield{author}{\bibinfo{person}{William~H. Walters}.} \bibinfo{year}{2023}\natexlab{}.
\newblock \showarticletitle{Assessing {{Diversity}} in {{Academic Library Book Collections}}: {{Diversity Audit Principles}} and {{Methods}}}.
\newblock \bibinfo{journal}{\emph{Open Information Science}} \bibinfo{volume}{7}, \bibinfo{number}{1} (\bibinfo{date}{Jan.} \bibinfo{year}{2023}).
\newblock
\showISSN{2451-1781}
\urldef\tempurl%
\url{https://doi.org/10.1515/opis-2022-0148}
\showDOI{\tempurl}


\bibitem[Watson(2019)]%
        {watsonHomosaurusDigitalTransgender2019}
\bibfield{author}{\bibinfo{person}{Brian Watson}.} \bibinfo{year}{2019}\natexlab{}.
\newblock \showarticletitle{Homosaurus and {{The Digital Transgender Archive}}}.
\newblock \bibinfo{journal}{\emph{The American Archivist Reviews Portal}} (\bibinfo{date}{June} \bibinfo{year}{2019}).
\newblock


\bibitem[Wyatt(2022)]%
        {wyattCollectionRebalance2022a}
\bibfield{author}{\bibinfo{person}{Neal Wyatt}.} \bibinfo{year}{2022}\natexlab{}.
\newblock \bibinfo{booktitle}{\emph{Collection {{Rebalance}} | 2022 {{Materials Survey}}}}.
\newblock Library Journal.
\newblock
\urldef\tempurl%
\url{https://www.libraryjournal.com/story/Collection-Rebalance-2022-Materials-Survey}
\showURL{%
\tempurl}
\newblock
\shownote{Accessed: 2025-01-22}.


\bibitem[Zwaaf(2020)]%
        {zwaafHomosaurusHttpHomosaurusorg2020a}
\bibfield{author}{\bibinfo{person}{Katrina Zwaaf}.} \bibinfo{year}{2020}\natexlab{}.
\newblock \showarticletitle{The {{Homosaurus}} {{http://homosaurus.org}}}.
\newblock \bibinfo{journal}{\emph{Technical Services Quarterly}} \bibinfo{volume}{37}, \bibinfo{number}{2} (\bibinfo{date}{April} \bibinfo{year}{2020}), \bibinfo{pages}{207--208}.
\newblock
\showISSN{0731-7131}
\urldef\tempurl%
\url{https://doi.org/10.1080/07317131.2020.1728140}
\showDOI{\tempurl}


\end{thebibliography}

\clearpage
\appendix
\section{APPENDIX}

\subsection{Interviews}\label{appendix-interviews}

We interviewed 14 staff members from public libraries across the United States, including collection development librarians, library directors, and library assistants. 
We conducted five initial interviews that helped inform our later survey (two in July 2023, three in late 2024–early 2025). 
Then, in January 2025, seven survey participants elected to participate in follow-up interview.
A recruitment email with a booking link to schedule a virtual meeting over a video conferencing software was sent to these survey respondents. 
Two interviewees were recruited via snowball sampling. 

\begin{table}[ht]
\centering

\resizebox{1\columnwidth}{!}{
\centering
\begin{tabular}{p{.5cm} p{2.5cm} p{2.5cm} p{5cm}} % set your desired widths
\toprule
\textbf{ID} & \textbf{US State or Tribal Nation} & \textbf{Community Type} & \textbf{Diversity Audit Tool(s)} \\
\midrule
P1 & Alabama & Town, Fringe & In-house \\
P2 & Georgia & City, Mid-size & Ingram \\
P3 & Illinois & Suburb, Large & Ingram, In-house \\
P4 & Indiana & City, Small & Baker \& Taylor (collectionHQ), In-house \\
P5 & Louisiana & City, Large & Baker \& Taylor (collectionHQ), Ingram \\
P6 & Massachusetts & Suburb, Large & In-house \\
P7 & Massachusetts & Suburb, Large & In-house \\
P8 & Michigan & Suburb, Large & Baker \& Taylor (collectionHQ), \newline Ingram, In-house, Overdrive \\
P9 & Missouri & City, Mid-size & Baker \& Taylor (collectionHQ), \newline In-house, Overdrive \\
P10 & North Carolina & City, Large & Baker \& Taylor (collectionHQ), \newline In-house \\
P11 & Oregon & Rural, Remote & Ingram \\
P12 & Texas & City, Large; Rural, Distant & In-house \\
P13 & Washington & Rural, Remote & Has not conducted a diversity audit \\
P14 & Native American \newline Reservation & Rural, Remote & Follett \\
\bottomrule
\end{tabular}
}

\caption{\textbf{List of interviewees and participant IDs, with corresponding information about diversity audit tools and library communities.} ``Community Type'' describes the size of the service community of the library that the interviewee worked at at the time of conducting diversity audit, or at the time of the interview (in the case of P13, who did not conduct a diversity audit). Community size categories were drawn from the FY 2022 Institute of Museum and Library Services Public Libraries Survey\cite{PLS2022}.}
\label{interview-table}
\end{table}

% 16 respondents scheduled interviews. However, nine fraudulent survey respondents scheduled interviews seeking compensation, despite none being offered to interviewees.

See Table \ref{interview-table} for a list of interviewees. 
Our interview protocol was reviewed by the IRB and determined to be exempt.

% \begin{adjustbox}{width=\columnwidth,center}{
% \def\sym#1{\ifmmode^{#1}\else\(^{#1}\)\fi}

% \begin{tabular}{rr}
% \toprule
% \textbf{Codes} \\
% \midrule
% Adult collections & Interpretation of Results \\
% Audit Motivations & Library Users \\
% Budget & Results \\
% Collection Size & Risks \\
% Comparison to Manual Audit & Time and Convenience \\
% COVID-19 & Transparency \\
% Data Quality & Vendors \\
% Diversity Categories & Where They Learned about Audit \\
% Diversity in Workforce & Young Readers Collections \\
% External Stakeholders & \\

% \bottomrule
% \end{tabular}
% \label{codes}
% }
% \end{adjustbox}

\subsection{Interview Instrument}
We provide a \underline{\href{https://drive.google.com/file/d/1Py9RCiWIjAOnBuO74E3CYOWxmK3tOqlp/view?usp=sharing}{PDF copy of our interview instrument}}. 
The linked questions were used as prompts for interviews via Zoom, with additional
follow-up questions asked based on responses. 
Interviewees were first asked for their
consent to participate in the interview and to record the interview for the purposes of transcription and internal review.

\subsection{Survey}\label{appendix-survey}
 We developed an online survey about librarians' experiences with and perspectives on collection diversity audits, which we shared with relevant mailing lists and online communities. The survey was open from January 14 to January 21. 
 We offered a drawing where respondents could win one of three \$100 Visa gift cards. 
 On one evening during the survey window, we received a large number of fraudulent responses from the same IP address, which prompted us to put the survey behind a password. 
 We used Qualtrics to process survey responses, and we filtered out the fraudulent responses and some other spam by cross-referencing IP addresses, timestamps, and the platform's fraud detection metrics.
 Responses were also reviewed for coherence and relevance to the survey topic.
 While efforts were made to identify and exclude low-quality responses, some fraudulent entries may have gone undetected. 
 This represents a potential limitation of our study.
 Our survey protocol was reviewed by the IRB and determined to be exempt.

 \subsection{Survey Instrument}
We provide a \underline{\href{https://drive.google.com/file/d/1GZ7XBBAW7oQBhuGNTVFFFqTD0nRgp9aA/view?usp=sharing}{PDF copy of our survey instrument}}.

% \begin{table}[ht]
%\centering
%\caption{Interviewees}
% \label{tab:interviewees}
 %\resizebox{\textwidth}{!}{
 %\begin{tabular}{|c|c|c|c|}
 %\hline
 %\textbf{ID} & \textbf{State} & \textbf{Community Type} & \textbf{Audit Tool(s)} \\
 %\hline
 %1 & P1 & Alabama          & Town, Fringe            & In-house \\
 %2 & P2 & Georgia          & City, Mid-size          & Ingram iCurate \\
 %3  & Indiana          & City, Small             & In-house, Baker \& Taylor Diversity Analysis Tool \\
 %4  & Louisiana        & City, Large             & Baker \& Taylor Diversity Analysis Tool, Ingram iCurate \\
 %5  & Massachusetts    & Suburb, Large           & In-house \\
 %6  & Massachusetts    & Suburb, Large           & In-house \\
 %7  & Michigan         & Suburb, Large           & In-house, Baker \& Taylor Diversity Analysis Tool, Ingram iCurate, OverDrive Diversity Audit \\
 %8  & Missouri         & City, Mid-size          & In-house, Baker \& Taylor Diversity Analysis Tool, OverDrive Diversity Audit \\
 %9  & Oregon           & Rural, Remote           & Ingram iCurate \\
 %10 & Texas            & City, Large; Rural, Distant & In-house \\
 %11 & Washington       & Rural, Remote           & No diversity audit conducted \\
 %\hline
 %\end{tabular}
 %}
 %\end{table}

\subsection{Spending Data}\label{spending}

\begin{figure*}
\centering
  \begin{tikzpicture}[scale=0.35, xscale = 2.0]
    \input{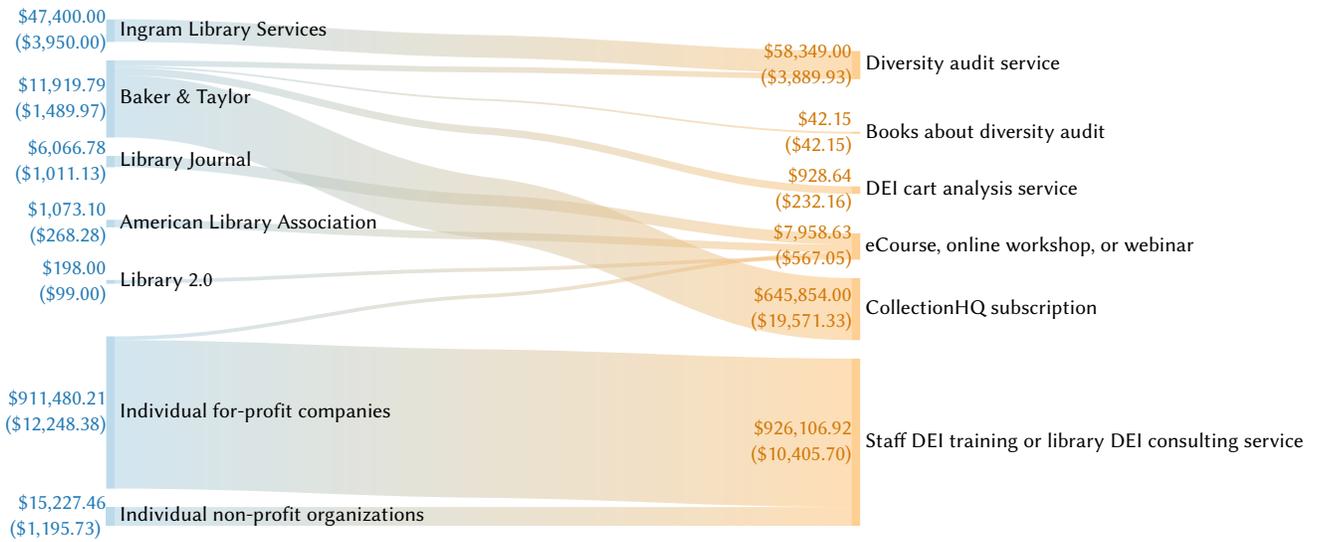}
  \end{tikzpicture}
  \caption{
  % Diversity audits cost several thousand dollars on average and cost libraries more than investing in the ability to conduct audits themselves (e.g. with the help of books about audits and webinars). 
  % Among the purchases we see on GovSpend, diversity audit services were provided by Ingram and Baker \& Taylor, and the price of such services cost several thousand dollars. A wider range of organizations provide services that aim to educate library employees on the concept of diversity audit and methods of in-house diversity audit. These services are mainly eCourse, online workshops, or webinars, and the cost of these services are much lower — around several hundred dollars. 
  % Note, we also see that libraries spend an average of \$10,000 on DEI training. And \$19,571 on Baker \& Taylor’s CollectionHQ. 
  This figure shows diversity-related purchases for 24 U.S. public libraries, drawn from GovSpend (for any item exceeding \$20 and including the words ``diversity'', ``dei'', ``icurate'', or ``collectionhq''). 
  Average item price is listed in parenthesis.
 Diversity audit services were provided by Ingram and Baker \& Taylor, and the price of such services cost several thousand dollars. 
  A wider range of organizations provide services that aim to educate library employees on the concept of diversity. 
  Libraries appear to spend more on staff DEI training and consulting than collection audits, averaging around \$10,000.
  However, as Table \ref{tab:LibA} and \ref{tab:LibB} show, libraries spend more with vendors that provide collection audits overall.
  % These services are mainly eCourse, online workshops, or webinars, and the cost of these services are much lower — around several hundred dollars. 
  % Purchase records are drawn from GovSpend and aggregated by type of seller and purchase; we include average item price in parenthesis.
  % These results reflect any purchases on GovSpend that include ``diversity'', ``dei'', ``icurate'', or ``collectionhq'' in the item description; include ``library'' in their agency name; exceed \$20; and were directly related to collection diversity audits. 
  % We search for purchases that include ``diversity'', ``dei'', ``icurate'', or ``collectionhq'', limit to agencies that include ``library'' in their name, and limit to purchases of \$20 or more.
  % These results include any purchases on GovSpend that include ``diversity'', ``dei'', ``icurate'', or ``collectionhq'' in the item description; include ``library'' in their agency name; and exceed \$20. 
  % These results represent general spending on diversity audit service or tools but exclude purchases of single titles. T
  % he filtering process yields 35 purchases from 24 libraries across 13 states. 
  % Reported figures represent the minimum amount libraries spent during this period.
  % 
  }
  \label{fig:sub1}
\end{figure*}

\begin{figure*}
\centering
  \begin{tikzpicture}[scale=0.65]
    \input{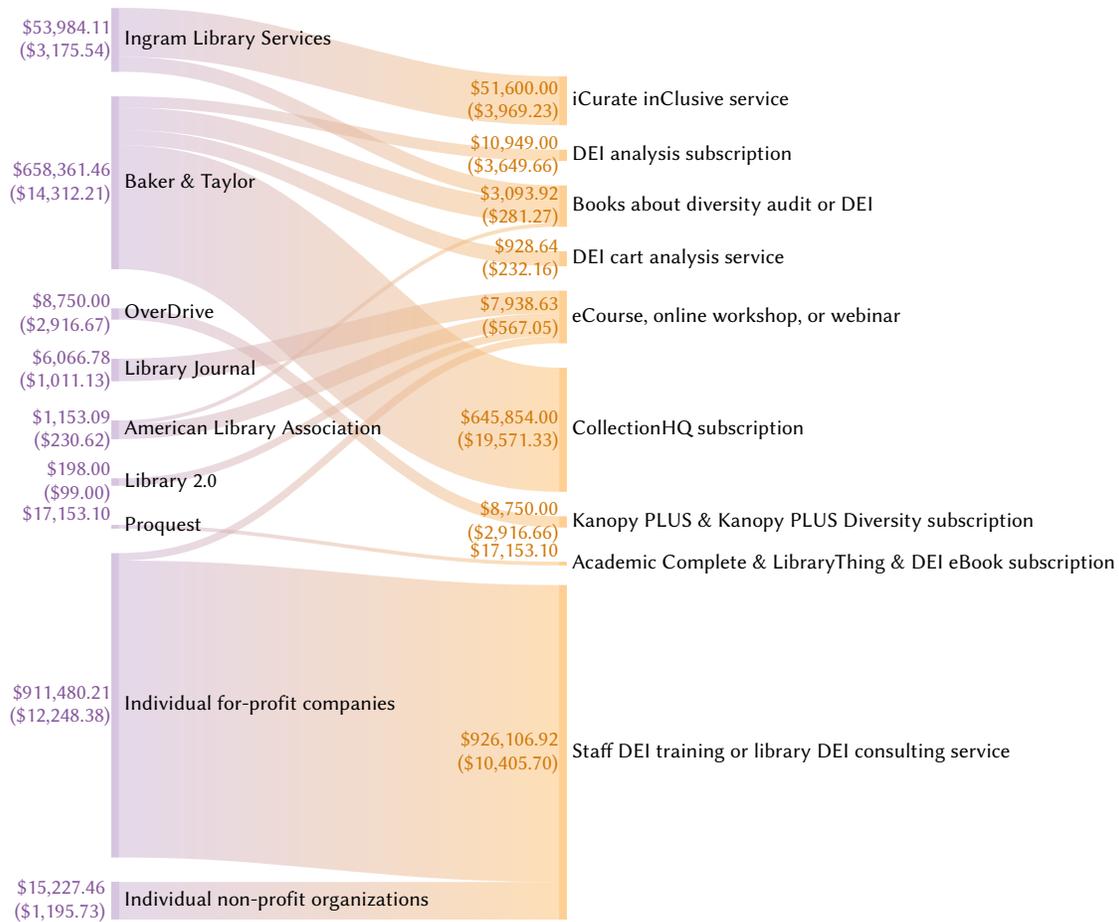}
  \end{tikzpicture}
  \caption{This figure shows diversity-related purchases from 57 U.S. public libraries. Purchase records are drawn from GovSpend and aggregated by type of seller and purchase; we include average item price in parenthesis.
  These results reflect any purchases on GovSpend that include ``diversity'', ``dei'', ``icurate'', or ``collectionhq'' in the item description; include ``library'' in their agency name; and exceed \$20. 
  % These results represent general spending on diversity but exclude purchases of single titles. 
  % The filtering process yields 171 purchases from 57 libraries across 21 states. 
These figures represent the minimum amount libraries spent during this period.}
  \label{fig:sub2}
\end{figure*}

We use GovSpend to explore public libraries' diversity-related purchases. We search for any item purchased by a public library that exceeds \$20 and includes the words ``diversity'', ``dei'', ``icurate'', or ``collectionhq''. The search results capture partial spending from 24 U.S. public libraries. We summarize and plot these data in Figure \ref{fig:sub1} and Figure \ref{fig:sub2}.

\end{document}